\documentclass[12pt, a4paper]{article}
\usepackage[dvipdfmx]{graphicx}
\usepackage{amssymb}
\usepackage{amsmath}
\usepackage{bm}
\usepackage{color}
\usepackage{theorem}
\usepackage{subcaption}
\usepackage{listings}
\usepackage{colortbl}
\usepackage{tabularx}
\usepackage{longtable}
\usepackage[T1]{fontenc}
\usepackage{lmodern}
\usepackage[top=30truemm,bottom=30truemm,left=25truemm,right=25truemm]{geometry}
\usepackage{multirow}
\usepackage[sort&compress,numbers, merge]{natbib}

\usepackage{braket}
\usepackage[compat=1.1.0]{tikz-feynman}
\usepackage{slashed}
\usepackage{comment}

\definecolor{Orange}{cmyk}{0,0.61,0.87,0}
\definecolor{JungleGreen}{cmyk}{0.99,0,0.52,0}
\definecolor{OliveGreen}{cmyk}{0.64,0,0.95,0.40}
\definecolor{Brown}{cmyk}{0,0.81,1,0.60}
\definecolor{RoyalBlue}{cmyk}{0.71,0.53,0,0.12}
\definecolor{Gray}{cmyk}{0,0,0,0.40}
\definecolor{LightPink}{cmyk}{0.0,0.25,0,0}
\definecolor{LLightPink}{cmyk}{0.0,0.10,0,0}
\definecolor{LightBlue}{cmyk}{0.25,0,0,0}
\definecolor{LightGray}{cmyk}{0,0,0,0.2}

\allowdisplaybreaks[1]

\usepackage[colorlinks=true, linkcolor=OliveGreen, citecolor=RoyalBlue,
urlcolor=RoyalBlue]{hyperref}

\renewcommand{\thefootnote}{\fnsymbol{footnote}}

\makeatletter
\newcommand\footnoteref[1]{\protected@xdef\@thefnmark{\ref{#1}}\@footnotemark}
\makeatother

\begin{document}

\begin{titlepage}

  \begin{flushright}
    {\tt
    }
\end{flushright}

\vskip 1.35cm
\begin{center}

{\Large 
{\bf
  Probing New Physics in the Vector-like Lepton \\[5pt] Model by Lepton Electric Dipole Moments
}
}

\vskip 1.5cm

Koichi~Hamaguchi$^{a,b}$\footnote{
\href{mailto:hama@hep-th.phys.s.u-tokyo.ac.jp}{\tt
 hama@hep-th.phys.s.u-tokyo.ac.jp}},
Natsumi Nagata$^a$\footnote{
\href{mailto:natsumi@hep-th.phys.s.u-tokyo.ac.jp}{\tt natsumi@hep-th.phys.s.u-tokyo.ac.jp}}, 
Genta Osaki$^a$\footnote{
\href{mailto:osaki@hep-th.phys.s.u-tokyo.ac.jp}{\tt osaki@hep-th.phys.s.u-tokyo.ac.jp }}, 
and 
Shih-Yen Tseng$^a$\footnote{
\href{mailto:shihyen@hep-th.phys.s.u-tokyo.ac.jp}{\tt shihyen@hep-th.phys.s.u-tokyo.ac.jp}}

\vskip 0.8cm

{\it $^a$Department of Physics, University of Tokyo, Bunkyo-ku, Tokyo
 113--0033, Japan} \\[2pt]
 {\it $^b$Kavli IPMU (WPI), University of Tokyo, Kashiwa, Chiba
  277--8583, Japan} \\[2pt]

\date{\today}

\vskip 1.5cm

\begin{abstract}
We examine the lepton dipole moments in an extension of the Standard Model (SM), which contains vector-like leptons that couple only to the second-generation SM leptons. The model naturally leads to sizable contributions to the muon $g-2$ and the muon electric dipole moment (EDM). One feature of this model is that a sizable electron EDM is also induced at the two-loop level due to the existence of new vector-like leptons in the loops. We find parameter regions that can explain the muon $g-2$ anomaly and are also consistent with the experimental constraints coming from the electron EDM and the Higgs decay $h\rightarrow \mu^{+}\mu^{-}$. The generated EDMs can be as large as $\mathcal{O}(10^{-22})~e \cdot \mathrm{cm}$ for the muon and $\mathcal{O}(10^{-30})~e \cdot \mathrm{cm}$ for the electron, respectively, which can be probed in future experiments for the EDM measurements.
\end{abstract}

\end{center}
\end{titlepage}

\renewcommand{\thefootnote}{\arabic{footnote}}
\setcounter{footnote}{0}

\section{Introduction}
Last year, Fermilab published their first result~\cite{Muong-2:2021ojo} on the measurement of the muon anomalous magnetic moment
\begin{equation}
  a_{\mu}\equiv \frac{g_{\mu}-2}{2} ~,
\end{equation}
which gives a value of
\begin{equation}
  a_{\mu}(\mathrm{exp}) = 116592061(41)\times 10^{-11} ~,
\end{equation}
while the Standard Model (SM) prediction is~\cite{Aoyama:2020ynm}
\begin{equation}
  a_{\mu}(\mathrm{SM}) = 116591810(43)\times 10^{-11} ~.
  \label{eq:delamu_SM}
\end{equation}
There is a $4.2\sigma$ tension between the experiment and theory,
\begin{equation}
  \Delta a_{\mu} = a_{\mu}(\mathrm{exp})-a_{\mu}(\mathrm{SM}) = 251(59)\times 10^{-11} ~,
\end{equation}
which may indicate the existence of physics beyond the Standard Model (BSM)\footnote{The hadronic vacuum polarization (HVP) contribution to the muon $g-2$ has been a challenge for theoretical calculations.
The value obtained in the data-driven method, which is adopted in Ref.~\cite{Aoyama:2020ynm}, is $a^{\mathrm{HVP}}_{\mu} = 6845(40) \times 10^{-11}$.
A discrepancy exists between this value and the lattice QCD calculations performed by the Budapest-Marseille-Wuppertal (BMW) group~\cite{Borsanyi:2020mff}, which gives $a^{\mathrm{HVP,LO}}_{\mu,\mathrm{BMW}} = 7075(55) \times 10^{-11}$. This is $2.1\sigma$ larger than the recommended data-driven result. Naturally, the two values should be compatible with each other because they correspond to the same physical processes in the SM. However, the current situation is that there is a significant difference between the two approaches, and the reason is not yet clear. More intriguingly, recent results from other lattice QCD groups~\cite{Ce:2022kxy,Alexandrou:2022amy} support the result obtained by the BMW group. For the time being, we simply fix on the value given in Eq.\,(\ref{eq:delamu_SM}).}.

Various new physics models have been proposed to explain this deviation. In general, these models contain hypothetical new particles and couplings, whose corresponding parameters are complex, and thus contain complex phases that break the $CP$ symmetry. It is well-known that the flavor-conserving $CP$ violation in the SM is very small, such that the induced particle electric dipole moments (EDMs) are vanishingly small.
The non-zero SM contributions to lepton EDMs appear at the four-loop level and are thus strongly suppressed. For example, the electron EDM is estimated to be $\left\vert d^{\mathrm{SM}}_{e} \right\vert \leq 10^{-38}~e \cdot \mathrm{cm}$~\cite{Pospelov:2005pr}.
Since it is far below the sensitivity of the current experimental techniques, any observation of a particle EDM will be an unambiguous sign of the new physics beyond the SM.

Currently, the upper bound on the muon EDM is
\begin{equation}
  \left\vert d_{\mu} \right\vert < 1.8 \times 10^{-19}~e \cdot \mathrm{cm}~\left(95\%~\mathrm{C.L.}\right)
\end{equation}
set by the Muon $\left(g-2\right)$ Collaboration at
Brookhaven National Laboratory~\cite{Muong-2:2008ebm}, which is about ten orders of magnitude weaker than the one on the electron EDM,
\begin{equation}
  \left\vert d_{e} \right\vert < 1.1 \times 10^{-29}~e \cdot \mathrm{cm}~\left(90\%~\mathrm{C.L.}\right)
\end{equation}
set by the ACME Collaboration~\cite{ACME:2018yjb}.
In order to improve the sensitivity on the muon EDM, there are several future experiments proposed to measure the muon EDM. For example, J-PARC Muon $g-2$/EDM experiment~\cite{Abe:2019thb} and the one using the frozen-spin technique at the Paul Scherrer Institute (PSI)~\cite{Adelmann:2021udj} will have sensitivities of $\sigma\left(d_{\mu}\right) \leq 1.5 \times 10^{-21}~e \cdot \mathrm{cm}$ and $\sigma\left(d_{\mu}\right) \leq 6 \times 10^{-23}~e \cdot \mathrm{cm}$, respectively.

In this paper, we consider a model with extra vector-like leptons (VLLs) as a possible explanation of the muon $g-2$ deviation, and investigate the EDMs of the muon and electron in the model. Models with VLLs have been discussed previously as solutions to the muon $g-2$ anomaly~\cite{Kannike:2011ng, Falkowski:2013jya, Dermisek:2013gta, Arnan:2016cpy, Megias:2017dzd, Kowalska:2017iqv, Raby:2017igl, Poh:2017tfo, Kawamura:2019rth, Hiller:2019mou, Hiller:2020fbu, Endo:2020tkb, Frank:2020smf, Chun:2020uzw, Kowalska:2020zve,Dermisek:2021ajd, Lee:2022nqz, deGiorgi:2022xhr}.
See also Refs.~\cite{Babu:2000dq, Ibrahim:2001jz, Feng:2001sq, Romanino:2001zf, Ellis:2001xt, Ellis:2001yza, Bartl:2003ju, Cheung:2009fc, Hiller:2010ib, Cesarotti:2018huy, Dekens:2018bci, Crivellin:2018qmi, Altmannshofer:2020ywf, Bigaran:2021kmn, Omura:2015xcg, Hou:2021zqq, Nakai:2022vgp, Dermisek:2022aec} for the works that studied the muon EDMs in the models motivated by the muon $g-2$ anomaly.
We consider a simple extension to the SM with two vector-like leptons, one SU(2)$_{L}$ doublet and one SU(2)$_{L}$ singlet.
We show that the model can naturally induce sizable EDMs of the muon and the electron at the one-loop and two-loop levels, respectively, in the parameter regions motivated by the muon $g-2$; the predicted EDMs can be as large as $|d_{\mu}|\sim 10^{-22}~e \cdot \mathrm{cm}$ and $|d_e|\sim 10^{-30}~e \cdot \mathrm{cm}$, which are within the reach of the proposed EDM experiments.
There are also researches discussing the indirect constraints on the muon EDM extracted from the EDM measurements of heavy atoms and molecules; for example, see Ref.~\cite{Ema:2021jds}.

This paper is structured as follows. In section \ref{sec:model}, we describe the model used in the analysis. In section \ref{sec:one_loop_muon_dm}, we summarize the calculation of the one-loop contributions to the muon dipole moments. In section \ref{sec:eEDM}, the induced electron EDM in this model is discussed. The experimental constraints on this model are presented in section \ref{sec:constraint}, and the results are given in section \ref{sec:result}. Finally, we summarize the study in section \ref{sec:summary}.


\section{Model}\label{sec:model}
\begin{table}
\centering
\begin{tabular}{l c c c | c c}
  \hline
    \hline
   & $\ell_L$ & $\mu_R$ & $H$ & $L_{L,R}$ & $E_{L,R}$ \\
 \hline
SU(3)$_{C}$ & {\bf 1} &  {\bf 1} & {\bf 1} & {\bf 1} & {\bf 1} \\ \hline
SU(2)$_{L}$ & {\bf 2} & {\bf 1} & {\bf 2} & {\bf 2} & {\bf 1} \\ \hline
U(1)$_{Y}$   & $-\frac{1}{2}$ & $-1$ & $\frac{1}{2}$ & $-\frac{1}{2}$ & $-1$ \\
   \hline
  \hline
      \end{tabular}
\caption{The quantum numbers of the SM and extra vector-like leptons in our model.} 
\label{tab:charges}
\end{table}
We consider an extension of the SM with one SU(2)$_L$ 
doublet ($L$) and one SU(2)$_L$ singlet ($E$) vector-like leptons. For simplicity, we consider the minimal scenario where the vector-like leptons couple only to the second-generation lepton, not to the first- and third-generation leptons.\footnote{Such kind of structure may be realized by imposing flavor symmetries on the model. We discuss a specific example in appendix \ref{appendix:muon_only}.}
Therefore, the electron and tau do not mix with extra vector-like leptons, and their masses totally originate from the Higgs Yukawa couplings. The quantum numbers of the fields necessary for the analysis are listed in Table~\ref{tab:charges}, where $\ell_L$ and $\mu_R$ are the second-generation leptons in the SM and $L_{L,R}$ and $E_{L, R}$ denote the vector-like leptons, which respectively have the same quantum numbers  as can be seen from the table. 
In the discussion below, our notation basically follows the one used in Ref.~\cite{Dermisek:2013gta}.

The components of the doublet fields are labeled as
\begin{align}
\ell_{L}=\begin{pmatrix} \nu_{\mu}\\ \mu_{L} \end{pmatrix}, ~L_{L,R}=\begin{pmatrix} L^{0}_{L,R}\\ L^{-}_{L,R} \end{pmatrix}, ~H=\frac{1}{\sqrt{2}}\begin{pmatrix} 0\\ v+h \end{pmatrix},
\end{align}
where $v=246.22$ GeV is the vacuum expectation value of the Higgs field. In the rest of this paper, we denote the indices of muon and two vector-like leptons in the mass basis as $2,4,5$, respectively, and for the muon, we use the symbol $\mu$ and $2$ interchangeably.

Without loss of generality, we work in the basis where the Yukawa matrix of the leptons in the SM sector is already diagonal. The most relevant part of the Lagrangian is the Yukawa interactions among the muon and the vector-like leptons and the mass term of the vector-like leptons, which are given by\footnote{In general, the other mass terms $M'_L \bar{\ell}_L L_R$ and $M'_R \bar{E}_L \mu_R$ are also allowed, but they can be removed by field redefinitions.} 
\begin{align}\label{eq:Lagrangian_VLL}
\begin{split}
\mathcal{L} \supset& -y_{\mu}\bar{\ell}_{L}\mu_{R}H - \lambda_{E}\bar{\ell}_{L}E_{R}H - \lambda_{L}\bar{L}_{L}\mu_{R}H - \lambda\bar{L}_{L}E_{R}H - \bar{\lambda}H^{\dag}\bar{E}_{L}L_{R} \\
&- M_{L}\bar{L}_{L}L_{R} - M_{E}\bar{E}_{L}E_{R} + \mathrm{h.c.}.
\end{split}
\end{align}
The parameters $y_{\mu},\lambda_{E},\lambda_{L},\lambda,\bar{\lambda},M_{L}$ and $M_{E}$ are in general complex. However, most of the complex phases are not physical since they can be removed via field redefinitions. It turns out that in the model there are two independent complex phases
\begin{align}
    \phi_\lambda &= \arg \left( y_{\mu}\lambda^{\ast}_{L}\lambda^{\ast}_{E}\lambda \right) ~, \\ 
    \phi_{\bar{\lambda}} &= \arg \left( y_\mu \lambda_L^* \lambda_E^* M_{L}M_{E}\bar{\lambda}^* \right)~,
\end{align}
which are invariant under the phase rotation of the fields and therefore can serve as the sources of new $CP$ violation. 
In this work, we take the two $CP$ phases to be the phases of $\lambda$ and $\bar{\lambda}$ and set the other parameters to be real. 

After the Higgs field develops a vacuum expectation value, the leptons acquire masses and the mass matrix of charged leptons is given by
\begin{align}\label{eq:mass_matrix}
\bar{f}_{L} M f_{R}
=
( \bar \mu_{L}, \bar{L}^{-}_{L}, \bar{E}_{L} ) 
\begin{pmatrix}
 y_{\mu} v/\sqrt{2} & 0 &  \lambda_{E} v/\sqrt{2}\\
  \lambda_{L} v/\sqrt{2} & M_{L} &  \lambda v/\sqrt{2}\\
 0 & \bar{\lambda} v/\sqrt{2} & M_{E} 
\end{pmatrix}
\begin{pmatrix}
 \mu_{R} \\
 L^{-}_{R}\\
 E_{R}
\end{pmatrix},
\end{align}
where the leptons in the flavor eigenbasis are denoted collectively as $f_{L}=(\mu_{L},L^{-}_{L},E_{L})^{T}$ and $f_{R}=(\mu_{R},L^{-}_{R},E_{R})^{T}$.

We can diagonalize the mass matrix in Eq.\,(\ref{eq:mass_matrix}) by a bi-unitary transformation with two unitary matrices $U_{L}$ and $U_{R}$:
\begin{align}
U^\dagger_L
\begin{pmatrix}
 y_{\mu} v & 0 &  \lambda_{E} v\\
  \lambda_{L} v & M_{L} &  \lambda v\\
 0 & \bar{\lambda} v & M_{E} 
\end{pmatrix}
U_R 
 =  
 \begin{pmatrix}
m_2  & 0 &   0\\
 0 & m_{4} &  0\\
 0 & 0 & m_{5} \\
\end{pmatrix},
\label{eq:mass}
\end{align}
where the eigenmass $m_{2}$ is set to be the mass of muon, $m_2=m_{\mu}$, and the mass ordering is fixed as $m_{2} < m_{4} < m_{5}$. We also note that the mass of the neutral component of the doublet $L$, $L^{0}\equiv\nu_{4}$, solely originates from the mass term $-M_{L}\bar{L}_{L}L_{R}$ and hence its mass is determined by the value of $M_{L}$. 

\section{Muon dipole moments}\label{sec:one_loop_muon_dm}
In this section, we summarize the one-loop contributions to the dipole moments of the muon, which are induced by the diagrams of the Higgs, $Z$ boson and $W$ boson mediations, as shown in Fig.~\ref{fig:one_loop_con}. For the relevant interactions, see appendix~\ref{appendix:VLL_couplings}.

\begin{figure}
\centering
\includegraphics[width=1.8in] {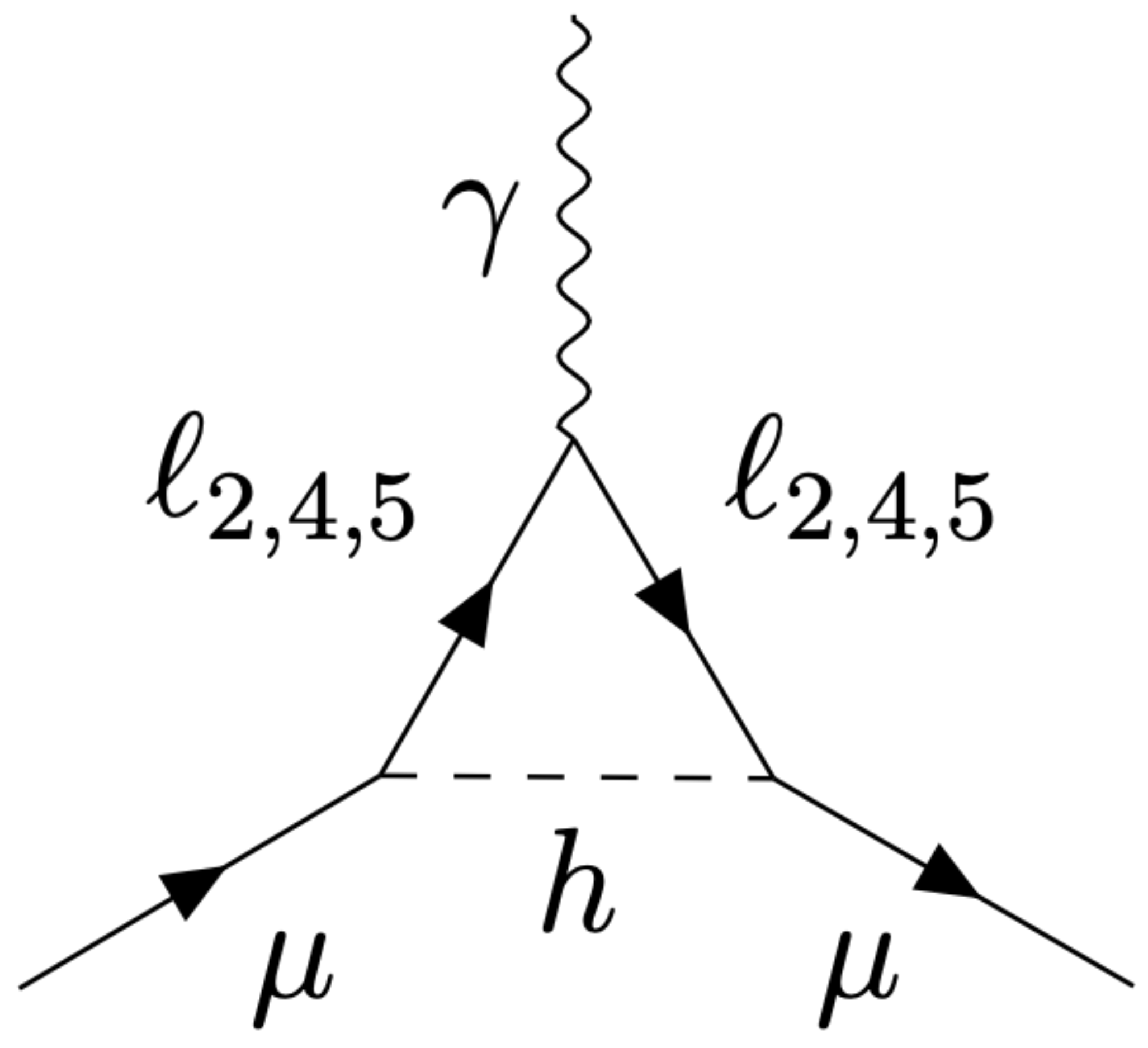}
\hspace{0.25cm}
\includegraphics[width=1.8in] {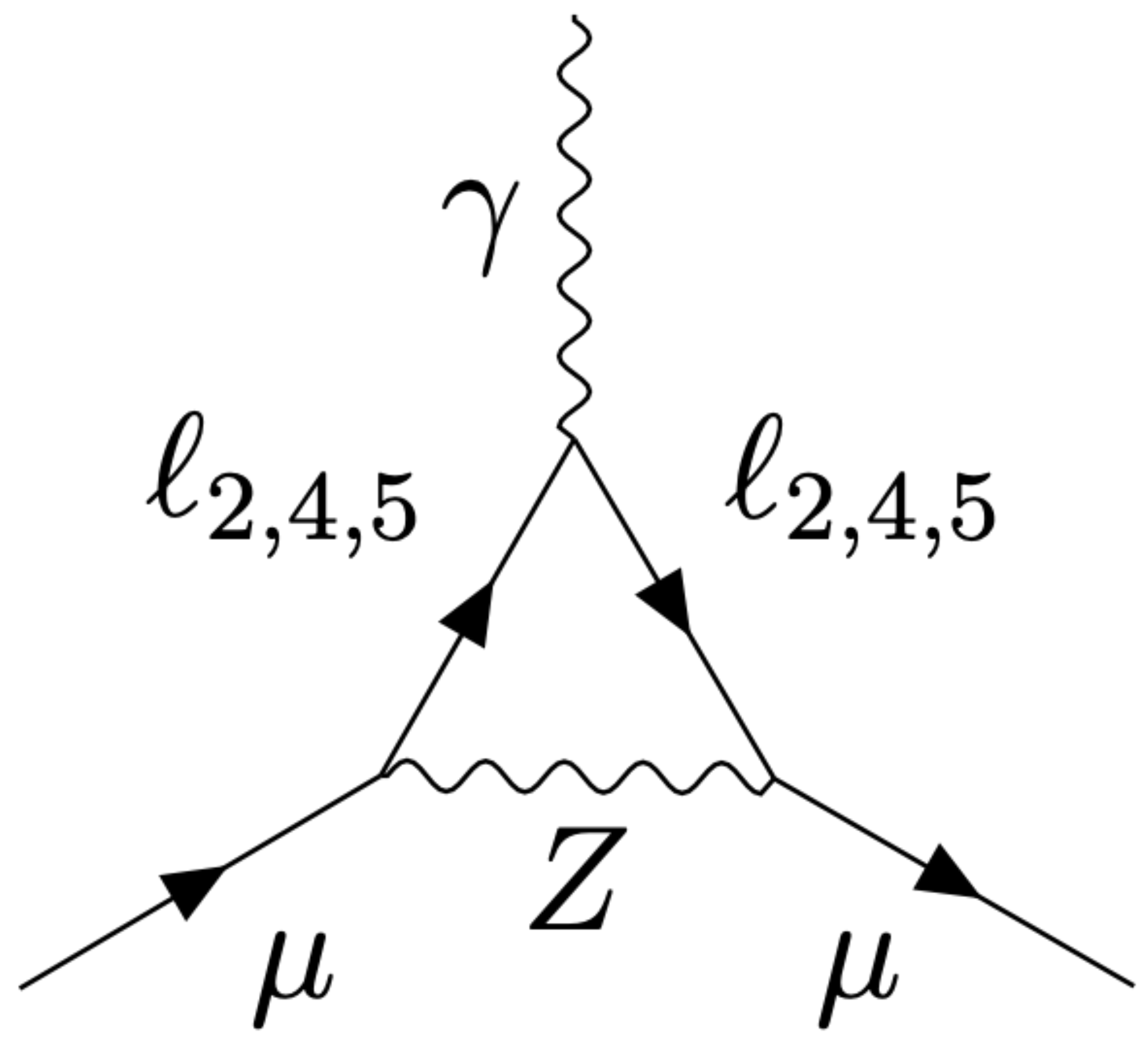}
\hspace{0.25cm}
\includegraphics[width=1.8in] {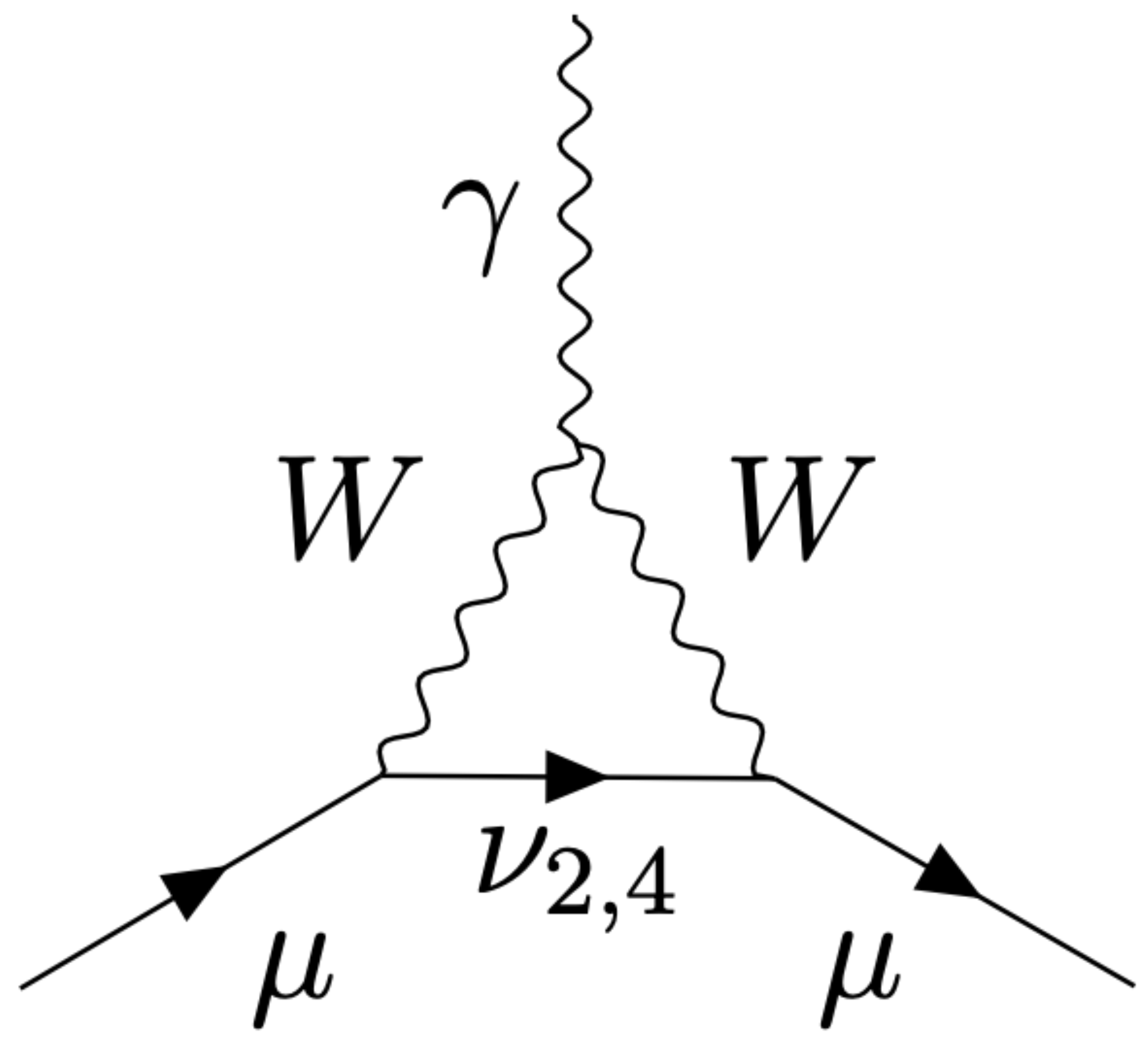}
\caption{One-loop contributions to the muon dipole moments. $\ell_{2,4,5}$ are the muon and two vector-like leptons in the mass basis, while $\nu_{2,4}$ are the muon neutrino and the heavy neutrino in the vector-like doublet, respectively.}
\label{fig:one_loop_con}
\end{figure}

\subsection{Higgs boson mediation}

The contributions to the moun dipole moments from the Higgs boson mediation are given by
\begin{align}
\Delta a_{\mu}^{h} &= \frac{m_\mu}{8\pi^2 m_h^2} \sum_{i=2,4,5}  \left[(|\lambda_{i2}|^2 + |\lambda_{2i}|^2) \, m_\mu f_h(r_{i}) +   {\rm Re}\, (\lambda_{i2}   \lambda_{2i} )  \, m_{i}  g_h(r_{i})  \right] - a^{h,\mathrm{SM}}_{\mu}, \label{eq:amu_h}\\
d_{\mu}^{h} &= -\frac{e}{16\pi^2 m_h^2} \sum_{i=2,4,5} {\rm Im} (\lambda_{i2} \lambda_{2i} ) m_{i} g_h(r_{i}),
\end{align}
where $a^{h,\mathrm{SM}}_{\mu}$ is the SM contribution from the diagram with muons in the loop. The loop functions are 
\begin{align}
g_{h}(r_{i}) &= -\frac{r_{i}^{2}-4r_{i}+3+2\,\mathrm{ln}(r_{i})}{2(1-r_{i})^{3}}, \\
f_{h}(r_{i}) &= \frac{r_{i}^{3}-6r_{i}^{2}+3r_{i}+2+6r_{i}\,\mathrm{ln}(r_{i})}{12(1-r_{i})^{4}},
\end{align}
with $r_{i}=m^{2}_{i}/m^{2}_{h}$ and $i=2,4,5$.

\subsection{$Z$ boson mediation}
The contributions to the moun dipole moments from the $Z$ boson mediation are given by
\begin{align}
\Delta a_\mu^Z &= \frac{m_\mu}{8 \pi^2 m_Z^2} \sum_{i=2,4,5}  \big\{  \left[|(g^{Z}_L)_{i2}|^2 + |(g^{Z}_R)_{i2}|^2\right] \, m_\mu  f_Z(r_{i})\notag \\
& \qquad \qquad \qquad \qquad + {\rm Re} \, \left[ (g^{Z}_L)_{i2} (g^{Z\ast}_R)_{i2} \right] \, m_{i}  g_Z(r_{i})  \big\} - a^{Z,\mathrm{SM}}_{\mu}, \label{eq:amu_Z}\\
d_{\mu}^{Z} &= \frac{e}{16\pi^2 m^{2}_{Z}} \sum_{i=2,4,5} {\rm Im} \, \left[ (g^{Z}_L)_{i2} (g^{Z\ast}_R)_{i2} \right] \, m_{i}  g_Z(r_{i}),
\end{align}
where $a^{Z,\mathrm{SM}}_{\mu}$ is the SM contribution from the diagram with muons in the loop. The loop functions are then given by
\begin{align}
g_{Z}(r_{i}) &= -\frac{r_{i}^{3}+3r_{i}-4-6r_{i}\,\mathrm{ln}(r_{i})}{2(1-r_{i})^{3}}, \\
f_{Z}(r_{i}) &= -\frac{5r_{i}^{4}-14r_{i}^{3}+39r^{2}_{i}-38r_{i}+8-18r^{2}_{i}\,\mathrm{ln}(r_{i})}{12(1-r_{i})^{4}},
\end{align}
with $r_{i}=m^{2}_{i}/m^{2}_{Z}$ and $i=2,4,5$.

\subsection{$W$ boson mediation}
The contributions to the muon dipole moments from the $W$ boson mediation are given by
\begin{align}
\Delta a^W_\mu &= 
\frac{m_\mu}{8 \pi^2 m_W^2} \big\{  \sum_{i=2,4}\left[ |(g^{W}_{L})_{i2}|^2 +|(g^{W}_R)_{i2}|^2\right] m_\mu  f_W(r_{i})\notag \\
& \quad \qquad \qquad \qquad + {\rm Re}\, \left[ (g^{W}_L)_{42} (g^{W\ast}_R)_{42} \right] \, M_{L} g_W(r_{4})\big\} - a^{W,\mathrm{SM}}_{\mu}, \label{eq:amu_W}\\
d^{W}_{\mu} &= \frac{e}{16\pi^2 m^{2}_{W}} {\rm Im}\, \left[ (g^{W}_L)_{42} (g^{W\ast}_R)_{42} \right] \, M_{L} g_W(r_{4}),
\end{align}
where $a^{W,\mathrm{SM}}_{\mu}$ is the SM contribution from the diagram with muon neutrino in the loop. The loop functions are given by
\begin{align}
g_{W}(r_{i}) &= \frac{r^{3}_{i}-12r^{2}_{i}+15r_{i}-4+6r^{2}_{i}\,\mathrm{ln}(r_{i})}{2(1-r_{i})^{3}}, \\
f_{W}(r_{i}) &= \frac{4r^{4}_{i}-49r^{3}_{i}+78r^{2}_{i}-43r_{i}+10+18r^{3}_{i}\,\mathrm{ln}(r_{i})}{12(1-r_{i})^{4}},
\end{align}
where 
$r_{2}=m^{2}_{\nu_{\mu}}/m^{2}_{W}$ assuming $m_{\nu_{\mu}} \simeq 0$, and $r_{4}=M^{2}_{L}/m^{2}_{W}$.

\section{Electron EDM}\label{sec:eEDM}
\begin{figure}
\center
\includegraphics[width=1.8in] {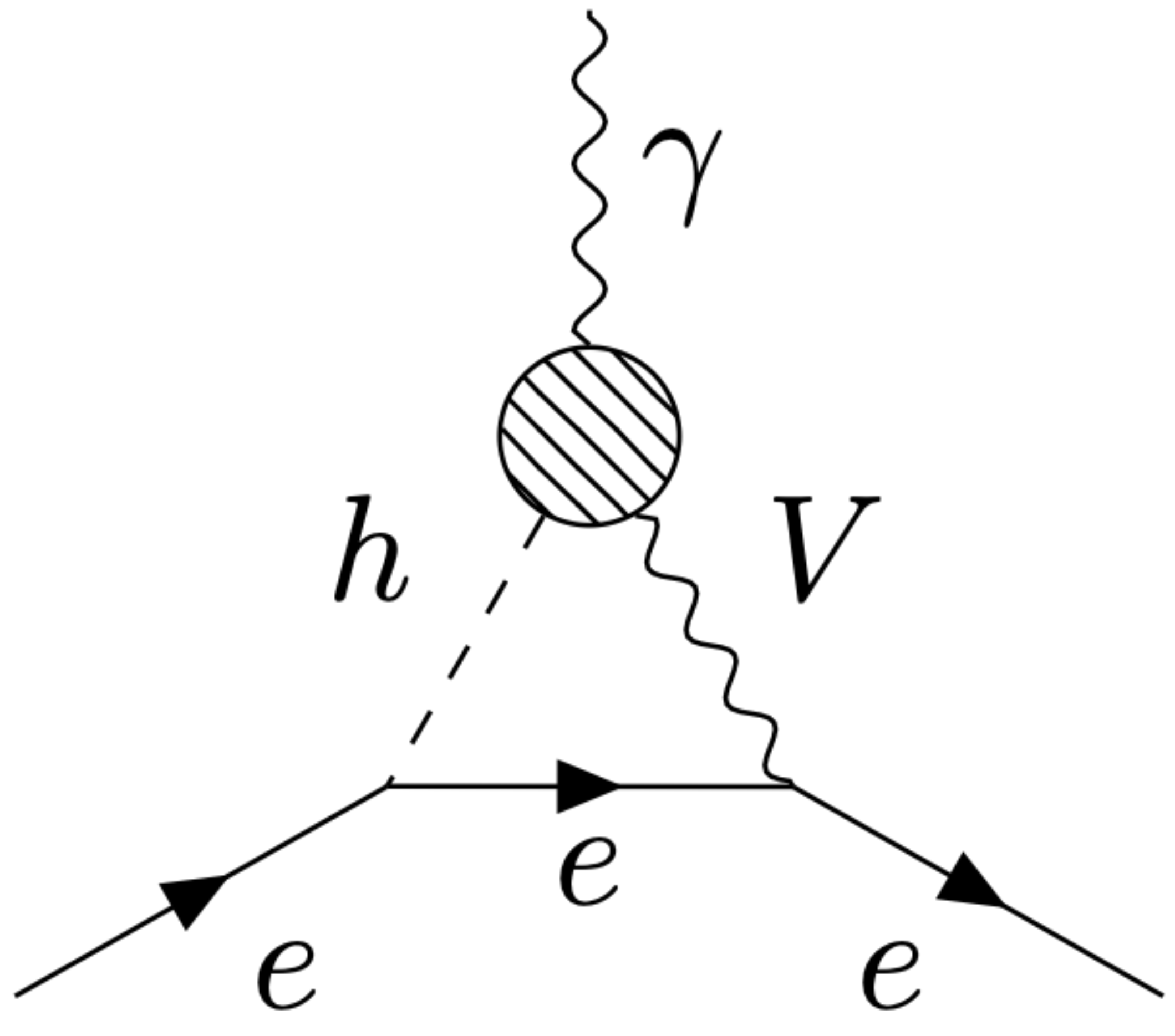}
\hspace{0.5cm}
\includegraphics[width=1.8in] {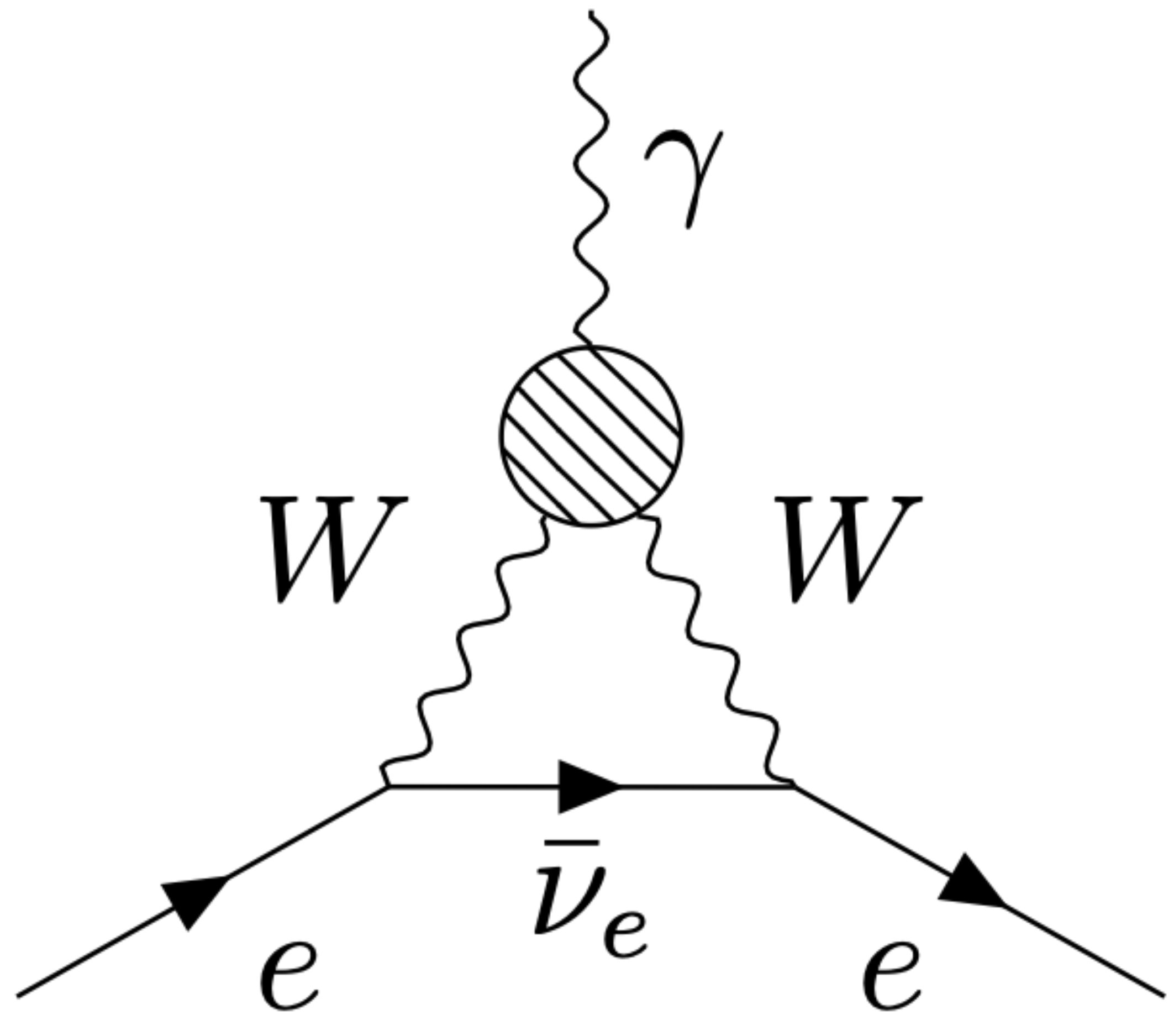}
\caption{Contributions to the electron EDM from the Barr-Zee diagram induced by the vector-like leptons.}
\label{fig:Barr_Zee}
\end{figure}

\begin{figure}
\center
\includegraphics[width=1.8in] {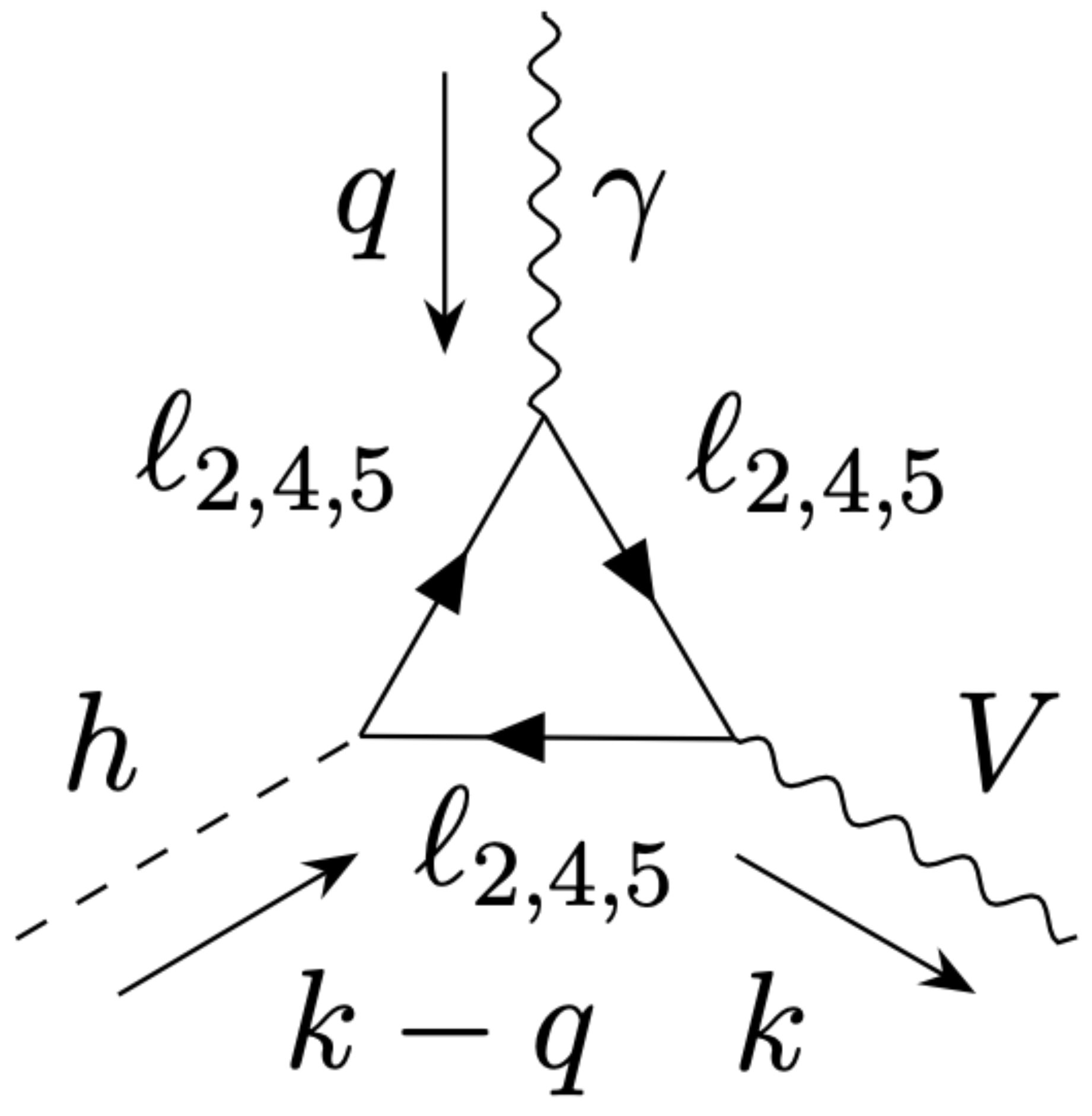}
\hspace{0.5cm}
\includegraphics[width=1.8in] {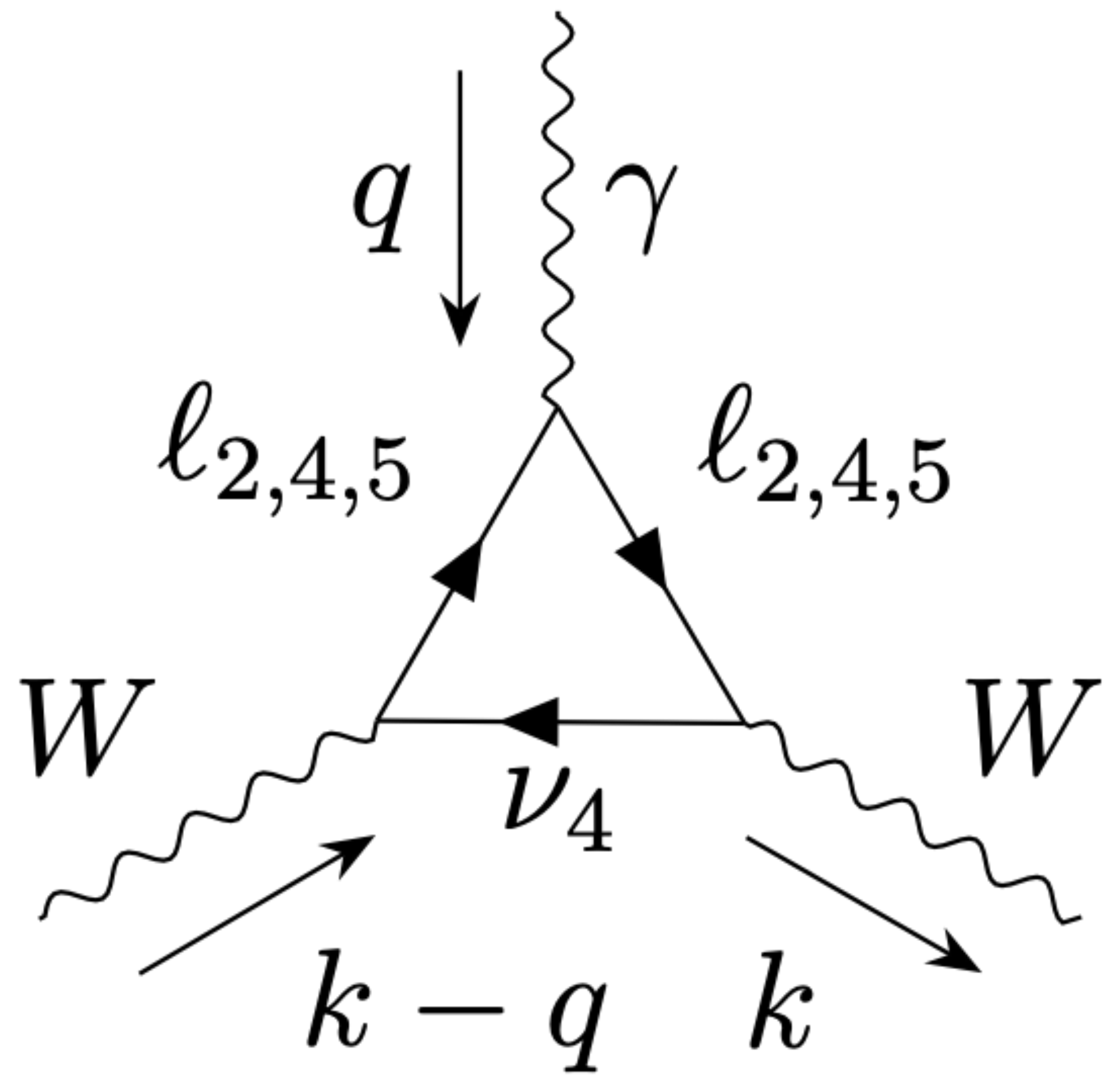}
\caption{The inner loop inside the blob of Fig.~\ref{fig:Barr_Zee}.}
\label{fig:Barr_Zee_inner}
\end{figure}

In this model, the mixings among the muon and the extra vector-like leptons also contribute to the electron EDM. These contributions appear in the internal lepton loop of the two-loop Barr-Zee diagrams~\cite{Barr:1990vd} shown in Fig.~\ref{fig:Barr_Zee}, where all possible combinations of muon and vector-like leptons contribute to the inner loop, as shown in Fig.~\ref{fig:Barr_Zee_inner}.

To obtain the value of the electron EDM, let us first discuss the left diagram in Fig.~\ref{fig:Barr_Zee}. The first step is to extract the effective vertex $\Gamma^{\mu\nu}_{hV\gamma}$ from the inner loop. By the gauge invariance of the on-shell photon, $q_{\mu}\Gamma^{\mu\nu}=0$, the effective vertex must have the form of~\cite{Nakai:2016atk}
\begin{align}
\Gamma^{\mu\nu,ij}_{hV\gamma}(q,k) = \int^{1}_{0} dx \frac{eQ}{4\pi^{2}D^{ij}} \left[ c_{E}^{Vij} (q^{\nu}k^{\mu} - q\cdot k\,g^{\mu\nu}) + ic_{O}^{Vij}\epsilon^{\mu\nu\alpha\beta}q_{\alpha}k_{\beta} \right],
\end{align}
where $V=\gamma$, $Z$, and $Q$ is the charge of the lepton coupled to the external photon which is $-1$ for all cases considered here. 
The coefficients $c_{E}^{Vij}$, $c_{O}^{Vij}$, and the $\Delta^{ij}$ in the denominator are given by
\begin{align}
c^{Vij}_{E} &= m_{i}x^{2}(1-x) (g^{ij\ast}_{s}g^{Vij}_{v} + g^{ij\ast}_{p}g^{Vij}_{a}) 
+ m_{j}(1-x)^{3} (g^{ij\ast}_{s}g^{Vij}_{v} - g^{ij\ast}_{p}g^{Vij}_{a}) , \\
c^{Vij}_{O} &= -m_{i}x(1-x) (g^{ij\ast}_{s}g^{Vij}_{a} + g^{ij\ast}_{p}g^{Vij}_{v}) 
+ m_{j}(1-x)^{2} (g^{ij\ast}_{s}g^{Vij}_{a} - g^{ij\ast}_{p}g^{Vij}_{v}) , \\
D^{ij} &= x(1-x)k^{2}-xm^{2}_{i}- (1-x)m^{2}_{j} .
\end{align}
Here, $i,j=2,4,5$, and $g_s,g_p,g_v^V,g_a^V$ are the couplings of scalar, pseudoscalar, vector, and axial-vector types, respectively.\footnote{\label{note} The details of interaction couplings can be found in appendix \ref{appendix:VLL_couplings}.}
The particle $j$ is defined as the one who interacts with the external photon, accompanied by particle $i$ in the inner loop of the Barr-Zee diagram.
The expression of the electron EDM from the $hV\gamma$ diagrams is given by
\begin{align}
\label{eq:electron_edm}
d^{hV\gamma}_e = 
\frac{eg^{Vee}_{v}g^{ee}_{s}}{32\pi^{4}} 
\sum_{i,j=2,4,5} \int^{1}_{0} dx \, \mathrm{Im}(c^{Vij}_{O}) \, I^{ij}_{hV},
\end{align}
where $g^{Vee}_{v}$ and $g^{ee}_{s}$ are the SM couplings of the electron to vector bosons, $V=\gamma$, $Z$, and Higgs boson\footnoteref{note}. The momentum integration over $k$ in the outer loop, $I^{ij}_{hV}$, is given by
\begin{align}
I^{ij}_{hV} = \frac{1}{m^{2}_{h}} \left[ F\left( x,\frac{m^{2}_{V}}{m^{2}_{h}},\frac{\Delta_{ij}}{m^{2}_{h}}\right) - F\left( x,\frac{m^{2}_{V}}{m^{2}_{h}},\frac{\Delta_{ij}}{m^{2}_{V}}\right) \right],
\end{align}
where 
\begin{align}
F(x,y,z) &= \frac{1}{(1-y)[z-x(1-x)]}
\mathrm{ln}\frac{z}{x(1-x)}, \\
\Delta_{ij} &= x m^{2}_{i} + (1-x) m^{2}_{j}.
\end{align}

Another class of diagrams is the ones with two $W$ bosons as shown in the right diagram of Fig.~\ref{fig:Barr_Zee}. In this case, the contribution to the electron EDM originates from the interaction of the heavy neutrino $\nu_{4}$ with the charged leptons, and is given by
\begin{align}
d^{WW\gamma}_{e} = \frac{eg^{2}}{256\pi^{4}} \sum_{i=2,4,5} \frac{m_{e}m_{i}M_{L}}{m^{2}_{W}} \mathrm{Im}\left[(g^{W}_{L})_{4i}(g^{W\ast}_{R})_{4i}\right] \int^{1}_{0} dx I^{4i}_{WW},
\end{align}
where
\begin{align}
I^{4i}_{WW} = \frac{(1-x)}{-x(1-x)m^{2}_{W}+(1-x)m^{2}_{i}+xM^{2}_{L}}
\mathrm{ln}\frac{(1-x)m^{2}_{i}+xM^{2}_{L}}{x(1-x)m^{2}_{W}}.
\end{align}

\section{Constraints}\label{sec:constraint}
In this section, we discuss the constraints on the model of vector-like leptons coming from the precision electroweak measurements, the electron EDM, the Higgs decay $h\rightarrow\mu^{+}\mu^{-}$, and the collider constraints on the heavy charged leptons.

\subsection{Precision electroweak measurements}
Since the muon and the muon neutrino mix with extra vector-like leptons in this model, the gauge couplings in the mass basis are modified (for details, see appendix \ref{appendix:VLL_couplings}). Therefore, several electroweak observables are affected correspondingly, such as the muon lifetime, decay asymmetries involving muons, partial widths of $W$ and $Z$ bosons, etc. These constraints have been considered in the previous analysis~\cite{Dermisek:2021ajd, Kannike:2011ng} and can be translated into upper bounds of the couplings $\lambda_{L}$ and $\lambda_{E}$ given by
\begin{align}\label{eq:lambda_constraint}
\frac{\lambda_{L}v}{\sqrt{2}M_{L}} \lesssim 0.04 ~,~ \frac{\lambda_{E}v}{\sqrt{2}M_{E}} \lesssim 0.03.
\end{align}

\subsection{\texorpdfstring{$h\rightarrow\mu^{+}\mu^{-}$}{Lg}}
Since the Yukawa coupling of the muon is also modified from its SM value, the deviation from the SM prediction is expected in the decay channel of the Higgs boson to a muon pair, $h\rightarrow \mu^{+}\mu^{-}$. 
In the latest search of this Higgs decay channel, the CMS group found the first evidence of the Higgs-to-dimuon decay channel~\cite{CMS:2020xwi} with a significance of three standard deviations. Currently, the branching fraction of $h\rightarrow \mu^{+}\mu^{-}$ is constrained to be within the range $0.8\times 10^{-4} < \mathcal{B}(h\rightarrow \mu^{+}\mu^{-}) < 4.5\times 10^{-4}$ at 95$\%$ confidence level. 
This can be transformed into
\begin{align}
0.37 < R(h\rightarrow \mu^{+}\mu^{-})\equiv\frac{\Gamma(h\rightarrow \mu^{+}\mu^{-})}{\Gamma(h\rightarrow \mu^{+}\mu^{-})_{\mathrm{SM}}} < 2.1,
\label{eq:h_to_mumu}
\end{align}
where we have used $\Gamma(h\rightarrow \mu^{+}\mu^{-})_{\mathrm{SM}}=2.16\times 10^{-4}$~\cite{LHCHiggsCrossSectionWorkingGroup:2016ypw} for $m_H=125.25$~GeV~\cite{Workman:2022}.
In our setup, the ratio is given by
$R(h\rightarrow \mu^{+}\mu^{-})=|\lambda_{22}|^2/(m_\mu/v)^2$.

\subsection{Electron EDM}
The latest measurement of the electron EDM is performed by the ACME collaboration~\cite{ACME:2018yjb}, which measured the precession of the electron spin in a superposition of the quantum states of an electron inside a strong intramolecular electric field. The group obtained an upper limit on the value of the electron EDM,
\begin{align}\label{eq:eEDM_constraint}
|d_{e}| < 1.1 \times 10^{-29}~e\cdot\mathrm{cm}
\end{align}
at 90$\%$ confidence level. 
We note that in obtaining the above limit, possible contributions to the spin precession frequency from the $CP$-odd electron-nucleon scalar coupling are set to zero. In the case of the model of vector-like leptons, such kinds of coupling exist as higher-order quantum effects and, therefore, can be safely ignored in our analysis.\footnote{The size of these higher-order effects can be estimated from the results in Ref.~\cite{Ema:2021jds}, where the contribution of the muon EDM to the electron EDM $d_e$ and the CP-odd electron-nucleon coupling $C_S$ is evaluated. This is expressed in terms of the equivalent electron EDM, $d_e^{\rm equiv}$, a linear combination of $d_e$ and $C_S$ that is constrained by experiments: $d_e^{\rm equiv} \simeq 5.8 \times 10^{-10} d_\mu$. As we see in the subsequent section, in our model we have $|d_\mu| \lesssim 1.5 \times 10^{-22}~e \cdot \mathrm{cm}$, and thus the size of the equivalent electron EDM induced radiatively by the muon EDM is $\lesssim 8.7 \times 10^{-32}~e \cdot \mathrm{cm}$, which is smaller than the Barr-Zee contribution for most of the parameter points shown in Fig.~\ref{fig:result_EDM}. }

\subsection{Direct search of heavy leptons}
The extra vector-like leptons decay via charged or neutral weak currents through the mixing with the SM second generation leptons. The tree-level decay modes include $\ell_i\to \ell_j + Z/h$, and $\ell_i\to \nu_j + W$, with $i>j$.
However, there are not so many experimental searches concentrating on vector-like leptons compared to the searches on vector-like quarks. 

The LEP experiment searched for such kinds of decays and set a lower bound on the mass of these new leptons~\cite{L3:2001xsz}. The lower bound was set to be around 100 GeV. Recently, the CMS group reported the lower bound on the mass of vector-like leptons coupled to the third generation of SM lepton, $\tau$~\cite{CMS:2022nty}. The doublet type is constrained to be heavier than 1045 GeV, and for the singlet type, the mass range 125--150 GeV is excluded. Expecting that the constraints on the vector-like leptons coupled to the muons are comparable, in this study, we assume that the vector-like leptons have masses of $\mathcal{O}(1)$~TeV.

\section{Results}\label{sec:result}
In this section, we show the results of our analysis on the model of vector-like leptons. 
We randomly choose the sampling points and scan over the parameters in the model with the ranges of parameters listed in Table~\ref{tab:parameters}. We note that the masses of the vector-like leptons are chosen to be at the TeV scale in this analysis so that the constraints from the direct search of heavy leptons are avoided. The muon Yukawa coupling $y_{\mu}$ is solved for the correct muon mass with all other 8 parameters fixed randomly in the range indicated by Table~\ref{tab:parameters}. We note that in solving the Yukawa coupling, it can be written in the form of a quadratic equation, therefore there are in general two solutions of Yukawa coupling for each randomly chosen parameter set. We include both of them in the result.
\begin{table}
\centering
\begin{tabular}{|c|c|}
  \hline
  Parameter  & Value \\
 \hline
   $M_{L}, M_{E}$ & $1 - 5$ TeV \\ \hline
   $|\lambda_{L}|, |\lambda_{E}|$ & $\leq$ EW constraints in Eq.\,(\ref{eq:lambda_constraint}) \\ \hline
   $|\lambda|, |\bar{\lambda}|$ & $0 - 1$ \\ \hline
   $\phi_{\lambda}, \phi_{\bar{\lambda}}$ & $0 - 2\pi$ \\ \hline
   $y_{\mu}$ & solved for the correct $m_{\mu}$ \\
  \hline
 \end{tabular}
\caption{Ranges of parameters randomly chosen for the sampling points. Around $4\times 10^{5}$ sampling points for the results are shown in the scattering plots.}
\label{tab:parameters}
\end{table}

\begin{figure}
\centering
\includegraphics[width=3.0in] {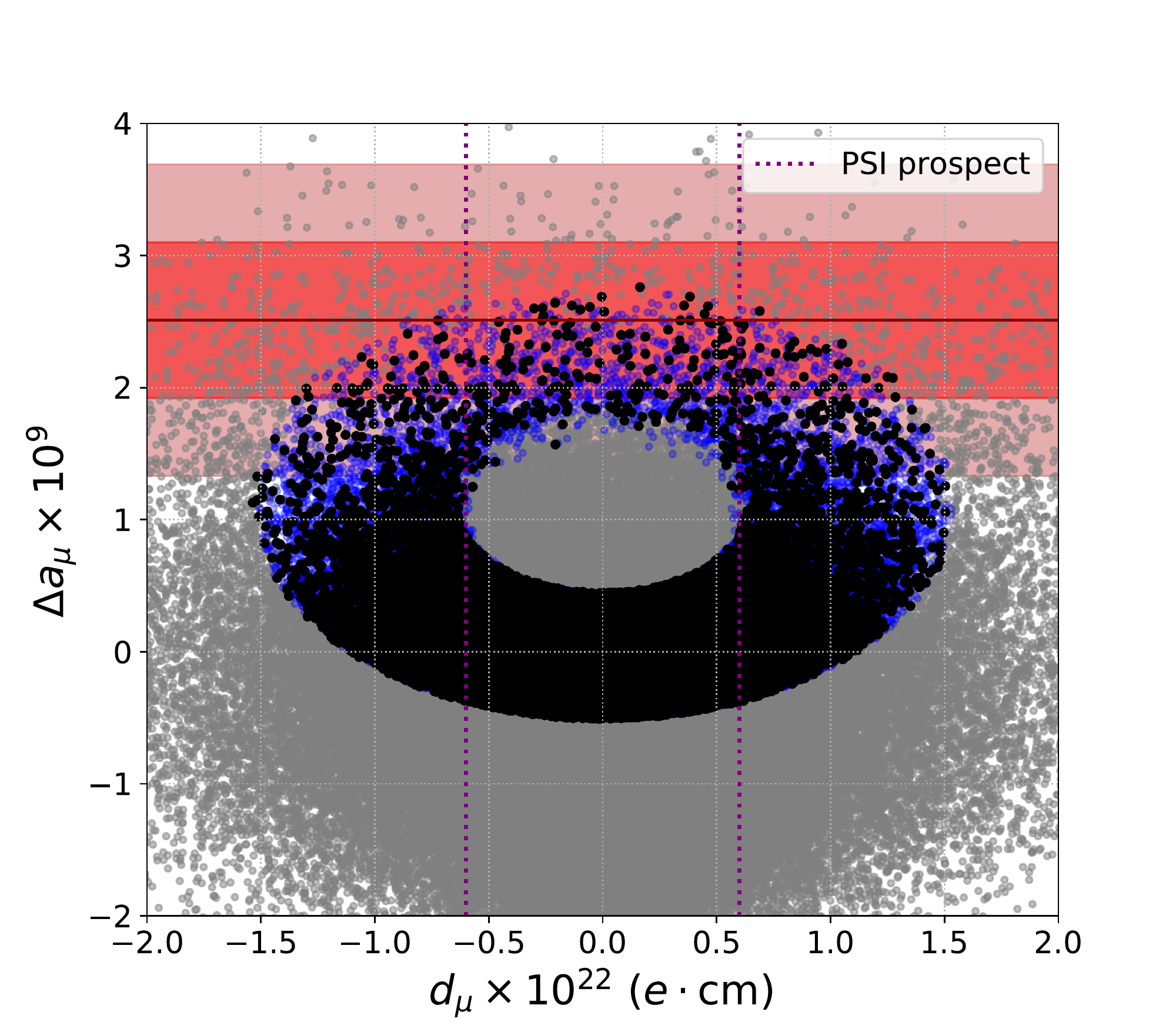}
\hspace{0.25cm}
\includegraphics[width=3.0in] {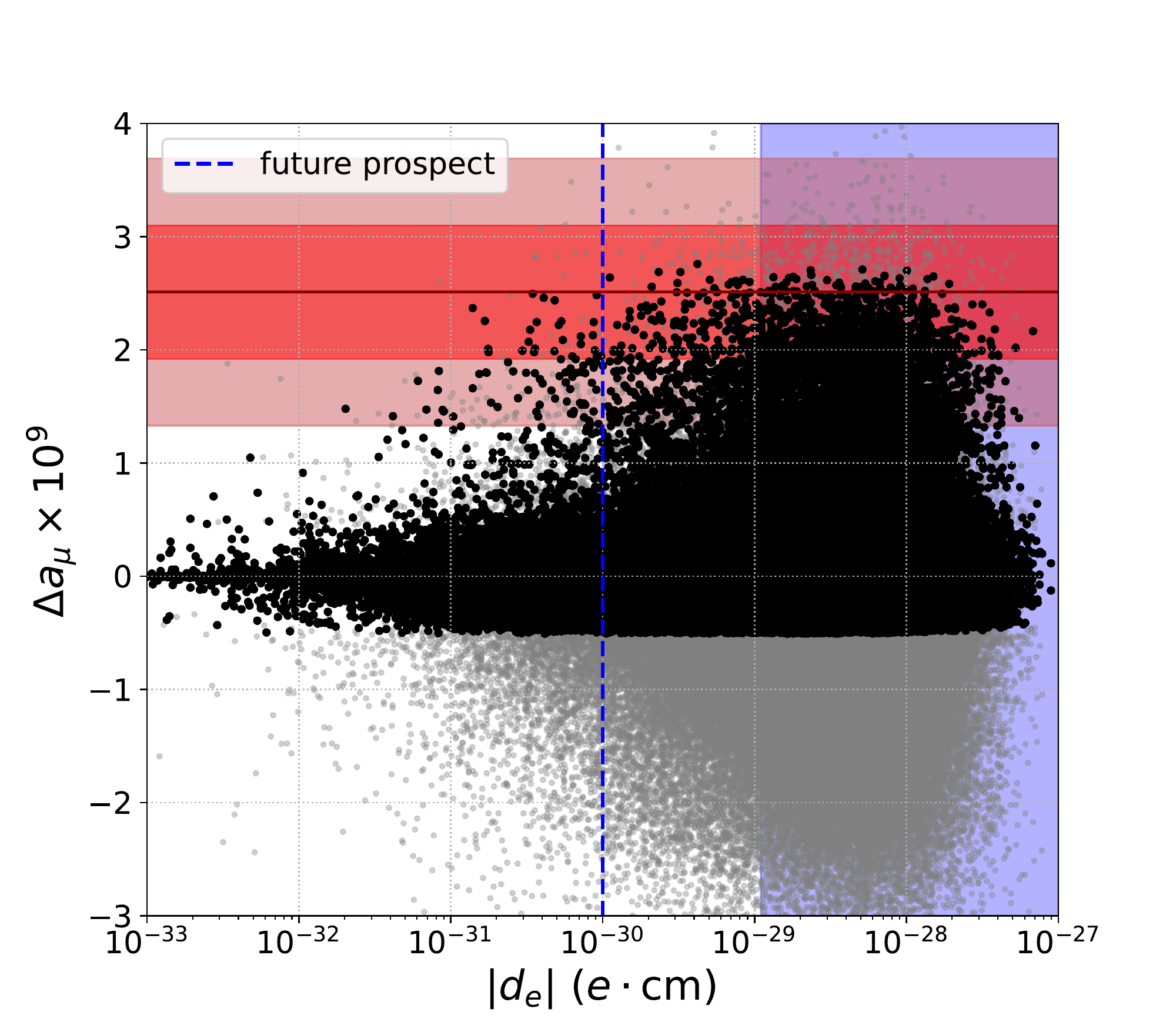}
\caption{The correlations of muon $g-2$ with muon EDM (left plot) and electron EDM (right plot). The horizontal red line and (light) red bands correspond to the central value and (2$\sigma$) 1$\sigma$ regions of the observed muon $g-2$, respectively. The $h\rightarrow \mu^{+}\mu^{-}$ constraint is first applied in both plots and the sampling points which do not satisfy it are shown as the gray dots. 
In the left plot, we also show the sampling points further excluded by the electron EDM constraint in blue.
The black dots are the points that satisfy both the constraints from $h\rightarrow \mu^{+}\mu^{-}$ and electron EDM. The vertical dotted line represents the PSI experiment.
In the right plot, the blue band shows the constraint from electron EDM measurement and the vertical dashed line indicates the prospect of the near-future electron EDM experiments.}
\label{fig:result_muon}
\end{figure}

In the left plot of Fig.~\ref{fig:result_muon}, the correlation between the muon $g-2$ and the muon EDM is shown. The horizontal red line and the (light) red bands correspond to the central value and (2$\sigma$) 1$\sigma$ regions of the observed muon $g-2$ in the experiment, respectively. The gray dots are the ones excluded by the $h\rightarrow \mu^{+}\mu^{-}$ constraint. After this selection, we further exclude the sampling points that predict the electron EDM to be too large and exceed the current upper bound given in Eq.\,(\ref{eq:eEDM_constraint}). These are represented by the blue dots. The black dots are the sampling points that satisfy both the constraints from $h\rightarrow \mu^{+}\mu^{-}$ and electron EDM. We can see that there are predictions consistent with the observed muon $g-2$, and the largest muon EDM is around $|d_{\mu}| \simeq 1.5 \times 10^{-22}~e \cdot \mathrm{cm}$. The PSI experiment~\cite{Adelmann:2021udj} can examine some regions of the parameters, and its sensitivity to the muon EDM is shown by the vertical dotted line. In this result, we confirm the elliptical correlation between muon $g-2$ and muon EDM obtained from the EFT approach~\cite{Dermisek:2022aec}. 

In the right plot of Fig.~\ref{fig:result_muon}, the correlation between the muon $g-2$ and the electron EDM is shown. The gray dots show the sampling points excluded by the $h\rightarrow \mu^{+}\mu^{-}$ constraint, and the region with the blue band is excluded by the current limit on the electron EDM from the ACME experiment. From the plot, we see that most of the sampling points that are consistent with both the $h\rightarrow \mu^{+}\mu^{-}$ and muon $g-2$ predict $|d_e|\gtrsim 10^{-32}~e \cdot \mathrm{cm}$.
The vertical dashed line, $|d_e|=10^{-30}~e\cdot \mathrm{cm}$, indicates a representative sensitivity of near future experiments. The next-generation ACME experiment (ACME III) is expected to improve the current sensitivity by an order of magnitude~\cite{ACMEexp}, and there are also proposals to improve the sensitivity by several orders of magnitude in the next 10-20 years~\cite{Alarcon:2022ero}.

\begin{figure}
\centering
\includegraphics[width=6.0in] {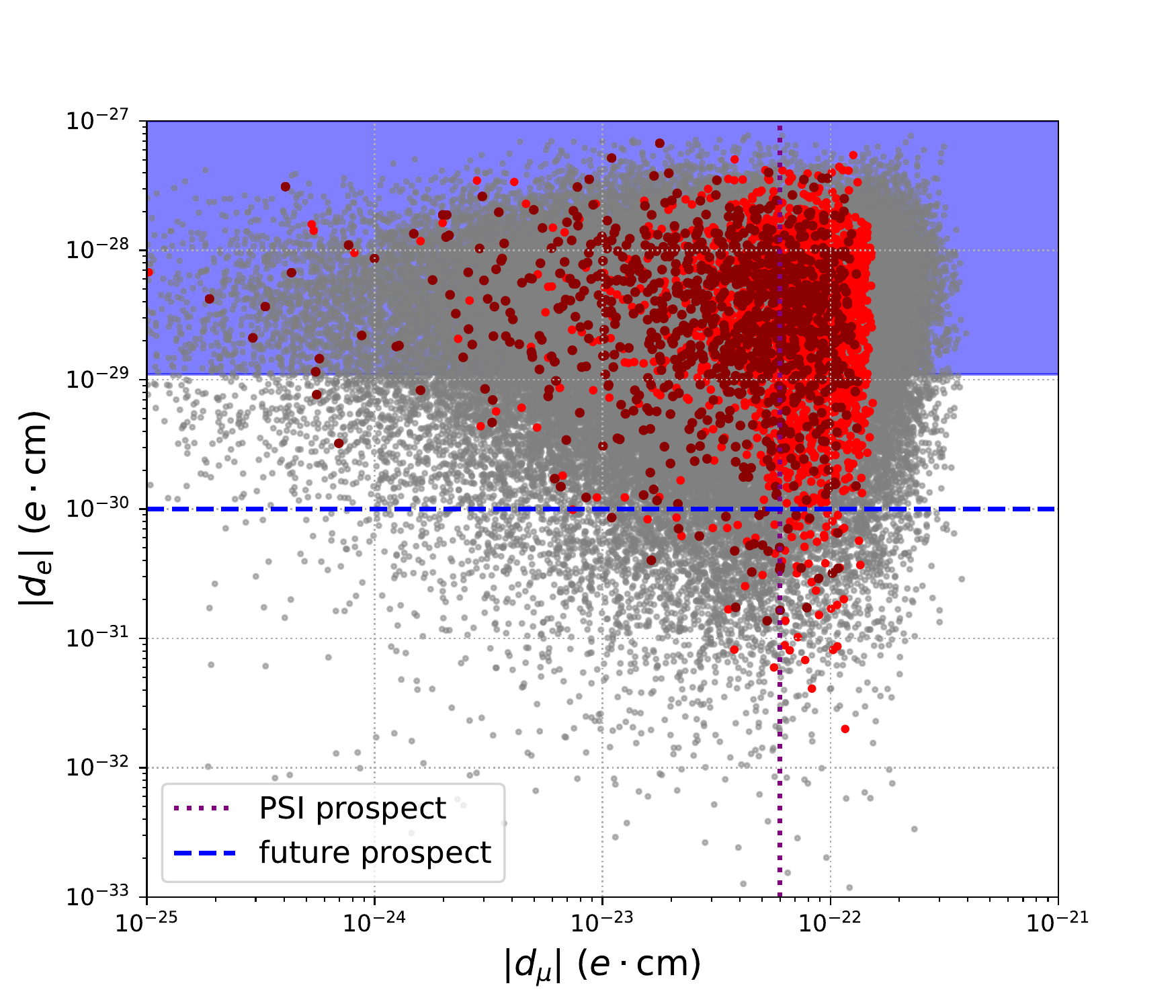}
\caption{The correlations between the muon and electron EDMs.
The gray dots show the points excluded by the $h\rightarrow \mu^{+}\mu^{-}$ constraint. The (dark) red dots represent the points satisfying the (1$\sigma$) 2$\sigma$ region of the muon $g-2$. The blue band is the constraint from the ACME experiment for the electron EDM measurement. The prospect of the PSI experiment for the muon EDM measurement and 
the near-future electron EDM experiments
are shown as the vertical dotted line and horizontal dashed line, respectively.}
\label{fig:result_EDM}
\end{figure}

In Fig.~\ref{fig:result_EDM}, we show the correlation between the muon and the electron EDMs. Again, we first excluded the sampling points that do not satisfy the $h\rightarrow \mu^{+}\mu^{-}$ constraint. These are the gray dots in the figure. The remaining sampling points are categorized into two classes, with the (dark) red dots representing the points satisfying the (1$\sigma$) 2$\sigma$ region of the muon $g-2$, respectively. The horizontal blue band is the constraint from the ACME experiment for the electron EDM measurement, and the prospects of the future muon/electron EDM experiments are also given in the plot. We find that these proposed experiments can investigate a significant fraction of the parameter space of this model, as they can exclude most of the available sampling points in the figure if no signals of EDMs are discovered. 
Furthermore, all of the allowed parameter space in Fig.~\ref{fig:result_EDM} can be covered by the electron EDM experiments offering sensitivities better than the current value by several orders of magnitude~\cite{Alarcon:2022ero}.

\begin{figure}
\centering
\includegraphics[width=3.0in] {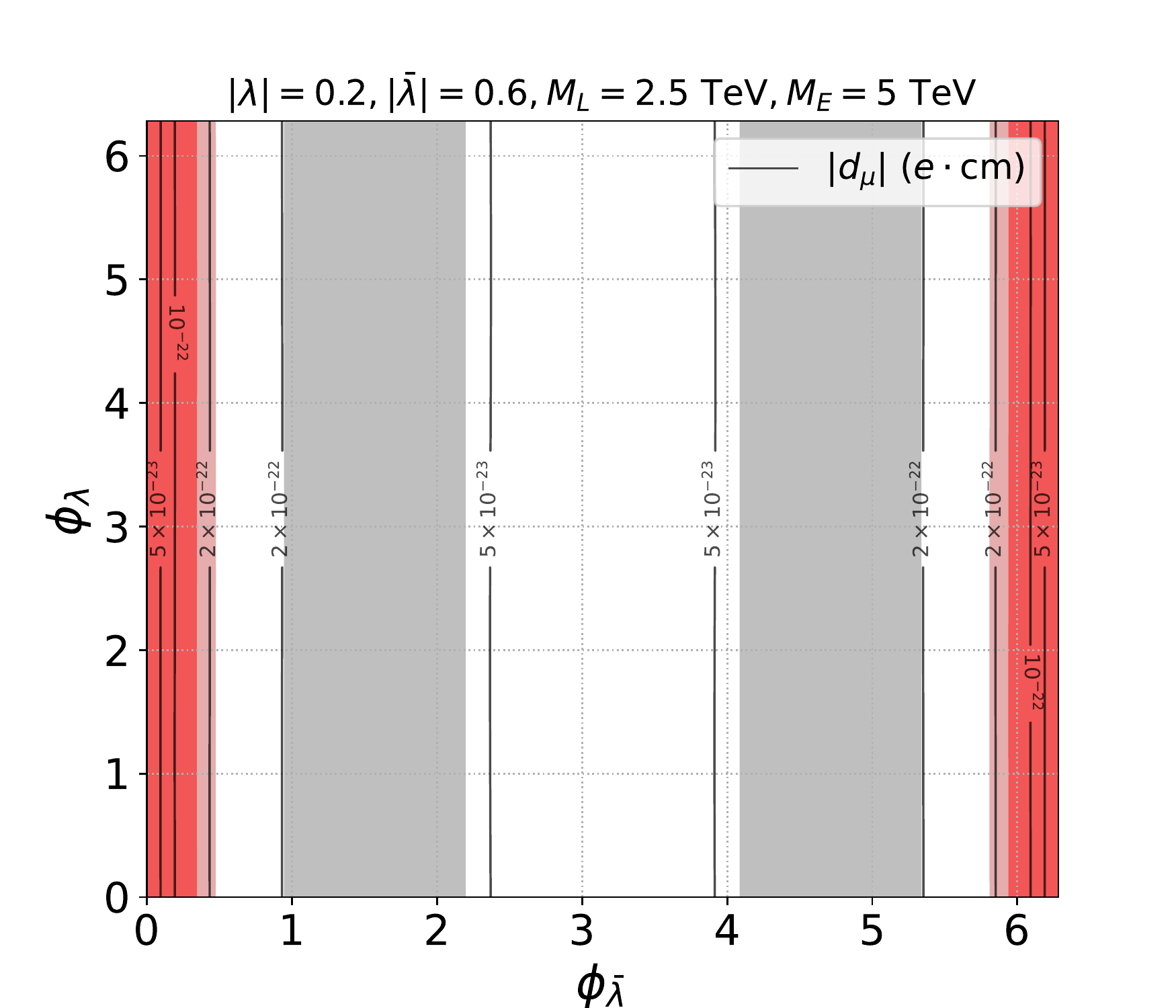}
\hspace{0.25cm}
\includegraphics[width=3.0in] {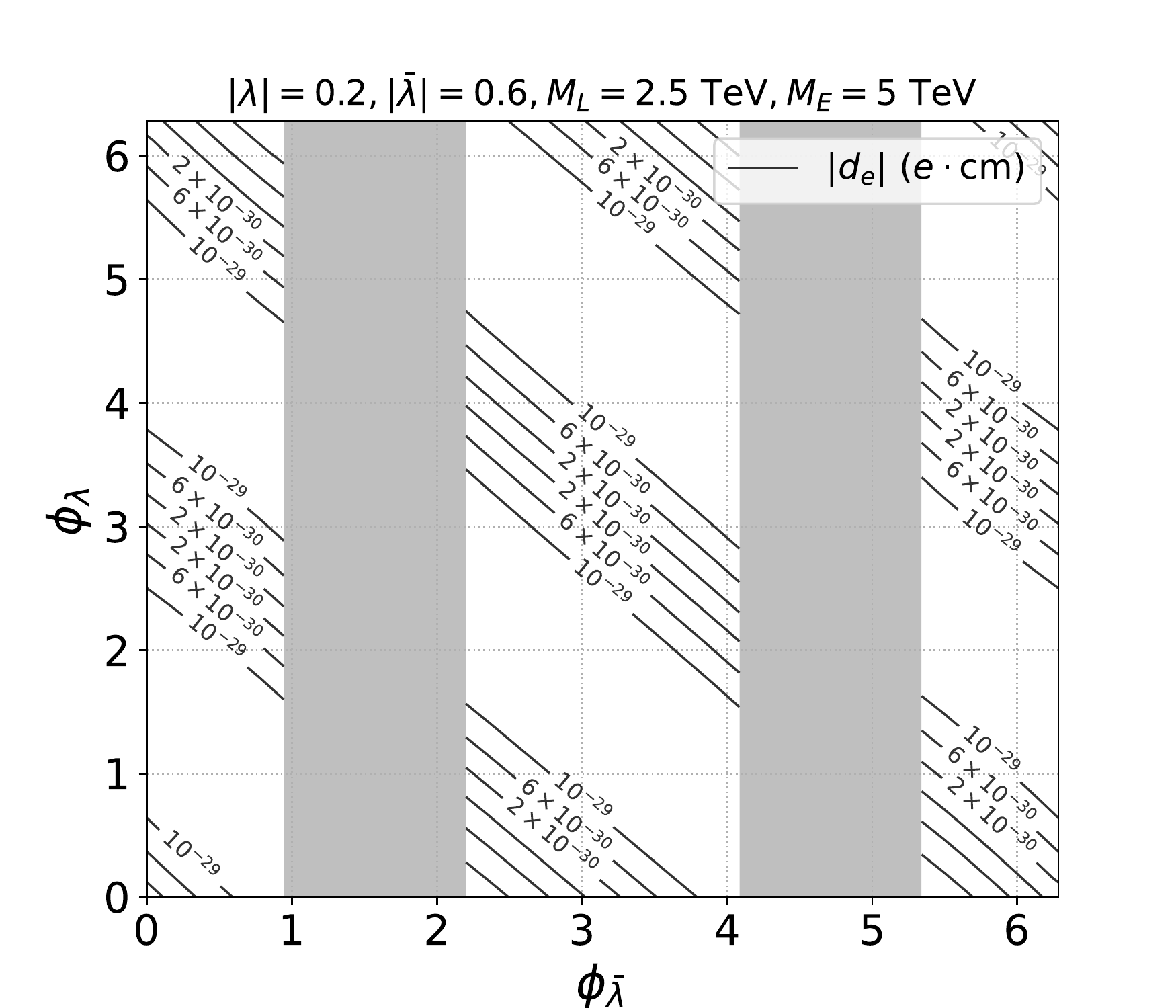}
\caption{The dependence of muon EDM (left) and electron EDM (right) on the two $CP$-violating phases in the model. The other parameters are chosen to be $|\lambda|=0.2,|\bar{\lambda}|=0.6,M_{E}=2.5~\mathrm{TeV},M_{L}=5~\mathrm{TeV}$, $\lambda_{L}=-0.04\times \sqrt{2}M_{L}/v\sim -0.57$, and $\lambda_{E}=0.03\times\sqrt{2}M_{E}/v\sim 0.86$. The size of the EDMs are given by the black contours. The gray bands represent the regions where no solution exists for the $y_\mu$ to give the correct muon mass after the mass diagonalization. The region consistent with muon $g-2$ at (2$\sigma$) 1$\sigma$ level is also indicated in the left figure by the (light) red bands. }
\label{fig:EDM_phase}
\end{figure}

Finally, we show the dependence of the muon and electron EDMs on the two $CP$-violating phases, $\phi_{\bar{\lambda}}$ and $\phi_{\lambda}$, in Fig.~\ref{fig:EDM_phase}. Here, we fix the other parameters to be $|\lambda|=0.2,|\bar{\lambda}|=0.6,M_{E}=2.5~\mathrm{TeV},M_{L}=5~\mathrm{TeV}$, $\lambda_{L}=-0.04\times \sqrt{2}M_{L}/v\sim -0.57$, and $\lambda_{E}=0.03\times\sqrt{2}M_{E}/v\sim 0.86$. The magnitude of the EDMs is indicated by the black contours. The gray bands are the regions where no solution exists for the $y_\mu$ to give the correct muon mass after the mass diagonalization.
The (2$\sigma$) 1$\sigma$ level of muon $g-2$ are also given in the left figure as the (light) red bands. As we can see from the plots, the muon EDM only depends on one of the two $CP$-violating phases, $\phi_{\bar{\lambda}}$, while in the case of the electron EDM, it depends on both of the phases. If the muon EDM is indeed nonzero and the vector-like leptons exist, we may be able to determine the masses and couplings of the vector-like leptons from the measurement of their decays, and the value of $\phi_{\bar{\lambda}}$ can then be fixed by the result of the muon EDM measurement.
We can input these parameters to the electron EDM and find the value of the remaining phase $\phi_{\lambda}$ according to the result of the electron EDM measurement.

\section{Summary}\label{sec:summary}
In this work, we have investigated a simple extension of the Standard Model, with the addition of one SU(2)$_L$ doublet and one SU(2)$_L$ singlet vector-like leptons, which are coupled to the second-generation SM leptons only. In this framework, the muon dipole moments are generated through the mediation of the Higgs, $W$, and $Z$ bosons. Furthermore, one interesting feature of this model is that a sizable electron EDM can also be induced at the two-loop level. Besides the latest value of the muon $g-2$ published by the Muon $g-2$ collaboration at the Fermilab, we have also considered the recent constraint on the Higgs decay channel $h\rightarrow \mu^{+}\mu^{-}$ reported by the CMS group and the constraint on the electron EDM from the ACME experiment. We have found that there are parameter regions consistent with all these constraints.

Because of the chirality flip from the heavy vector-like leptons, the sizes of muon dipole moments are enhanced. The muon EDM can be as large as $10^{-22}~e \cdot \mathrm{cm}$, which can be probed in the future muon EDM measurement such as the PSI experiment, whose sensitivity is estimated to be around $6\times 10^{-23}~e \cdot \mathrm{cm}$~\cite{Adelmann:2021udj}. An electron EDM of $\mathcal{O}(10^{-30})~e \cdot \mathrm{cm}$ can be generated in this model, and future electron EDM measurement like the ACME III experiment, whose sensitivity is expected to be one order of magnitude better than the previous ACME II~\cite{ACMEexp}, is able to examine the prediction.

\section*{Acknowledgments}

This work is supported in part by the Grant-in-Aid for Innovative Areas (No.19H05810 [KH], No.19H05802 [KH], No.18H05542 [NN]), Scientific Research B (No.20H01897 [KH and NN]), Young Scientists (No.21K13916 [NN]), and JSPS KAKENHI Grant (No.20J22214 [SYT]).

\appendix
\section{Interactions}\label{appendix:VLL_couplings}
In this appendix, we briefly summarize the interactions in the model we used in the analysis. Concerning the index notation, the quantities related to the flavor basis are indexed by Greek alphabets ($\alpha,\beta,\gamma,\cdots$), while the quantities related to the mass basis are indexed by Latin alphabets ($i,j,k,\cdots$). For the flavor indices, we have $2=\mu,4=L$ and $5=E$.
\subsection{Higgs boson couplings}
Since the couplings related to electron and tau are not modified by the vector-like leptons, their couplings with the Higgs boson are the SM values $\lambda_{e,\tau} = -m_{e,\tau}/v$. Other couplings of charged leptons to the Higgs boson are modified by the mixings.
The Yukawa interaction in the flavor basis is given by
\begin{eqnarray}
{\cal L}_{Y} \supset  - \frac{1}{\sqrt{2}} \, \bar f_{L\rho} \, Y_{\rho\sigma} \, f_{R\sigma} \, h + \mathrm{h.c.}.
\end{eqnarray}
This can be transformed to the the mass basis given by
\begin{align}
{\cal L}_{Y} \supset -\frac{1}{\sqrt{2}} \sum_{\rho,\sigma=2,4,5} \bar{f}_{Li} (U^{\dag}_{L})_{i\rho} \, Y_{\rho\sigma} \, (U_{R})_{\sigma j} \, f_{Rj} \, h + \mathrm{h.c.} \equiv \bar{f}_{Li} \, \lambda_{ij} \, f_{Rj} \, h + \mathrm{h.c.},
\end{align}
where the Yukawa matrix is written as
\begin{align}
Y = 
\begin{pmatrix}
 y_\mu  & 0 &  \lambda_{E} \\
  \lambda_{L}  &0 &  \lambda \\
 0 & \bar \lambda  & 0
\end{pmatrix},
\end{align}
and the Yukawa couplings in the mass basis is
\begin{align}
\lambda_{ij} =  -\frac{1}{\sqrt{2}} \sum_{\rho,\sigma=2,4,5}  (U^\dagger_L)_{i\rho} Y_{\rho\sigma} \left(U_R\right)_{\sigma j}.
\label{eq:lamab}
\end{align}
We notice that the Yukawa coupling in the flavor basis is similar to the mass matrix of the charged leptons with the absence of the masses of vector-like leptons, that is, $Y v/\sqrt{2} = M - \mathrm{diag} (0,M_L,M_E)$. With this relation, the Higgs boson couplings in the mass basis can be written as:
\begin{align}
-\lambda v &= 
U^\dagger_L
\begin{pmatrix}
 y_{\mu} v/\sqrt{2} & 0 &  \lambda_{E} v/\sqrt{2} \\
  \lambda_{L} v/\sqrt{2} & M_{L} &  \lambda v/\sqrt{2} \\
 0 & \bar{\lambda} v/\sqrt{2} & M_{E} 
\end{pmatrix} 
U_R
- U^\dagger_L
\begin{pmatrix}
 0  & 0 &  0 \\
  0  &M_L&  0 \\
 0 & 0  & M_E
\end{pmatrix}
U_R \\
&=
\begin{pmatrix}
 m_\mu  & 0 &  0 \\
  0  &m_{4}&  0 \\
 0 & 0  & m_{5}
\end{pmatrix} 
- U^\dagger_L
\begin{pmatrix}
 0  & 0 &  0 \\
  0  &M_L&  0 \\
 0 & 0  & M_E
\end{pmatrix} 
U_R.
\end{align}
In this expression, we can see clearly that the first term has a simple form which is the same as the Yukawa couplings in the SM. The second term corresponds solely to the contributions from the mass term of the vector-like leptons.

\subsection{$Z$ boson couplings}
The couplings of leptons to the $Z$ boson come from the kinetic term of the leptons. Due to the mixing among muon and vector-like leptons, the kinetic term is modified to
\begin{align}
{\cal L}_{\mathrm{kin}} \supset \bar f_{L\sigma} i\slashed{D}_{\sigma} f_{L\sigma} + \bar f_{R\sigma} i\slashed{D}_{\sigma}  f_{R\sigma} = \bar f_{Li} (U^{\dag}_{L})_{i \sigma} i\slashed{D}_{\sigma} (U_{L})_{\sigma j} f_{Lj} + \bar f_{Ri} (U^{\dag}_{R})_{i \sigma} i\slashed{D}_{\sigma} (U_{R})_{\sigma j}  f_{Rj}.
\end{align}
The covariant derivative $D_{\mu\sigma}$ is defined as
\begin{align}
D_{\mu\sigma} = \partial_{\mu} + i\,\frac{g}{\mathrm{cos}\,\theta_{W}}(T^{3}_{\sigma}-Q_{\sigma}\mathrm{sin}^{2}\theta_{W})Z_{\mu}+ieQ_{\sigma}A_{\mu},
\end{align}
where $e$ and $g$ are the electromagnetic and the SU(2) couplings of the Standard Model, and $T^{3}$ and $Q$ are the weak isospin and electric charge of leptons obtained from the quantum numbers listed in Table \ref{tab:charges}.

The couplings of the $Z$ boson to charged leptons $\ell_i$ and $\ell_j$ are defined in the Lagrangian
\begin{align}
{\cal L}_{Z} \supset \left[ \bar f_{Li} \gamma^\mu (g^{Z}_{L})_{ij}  f_{Lj} + \bar f_{Ri} \gamma^\mu (g^{Z}_{R})_{ij}  f_{Rj} \right] Z_\mu.
\end{align}
The couplings of left-handed and right-handed charged leptons with $Z$ boson are given by
 \begin{align}
(g^{Z}_{L})_{ij} = -\frac{g}{\cos \theta_W} \displaystyle\sum\limits_{\sigma=2,4,5} (T^3_{L\sigma} - \sin ^2 \theta_W Q_\sigma) (U^\dagger_L)_{i\sigma} (U_L)_{\sigma j}, 
\label{eq:gZL}\\
(g^{Z}_{R})_{ij} = -\frac{g}{\cos \theta_W} \displaystyle\sum\limits_{\sigma=2,4,5} (T^3_{R\sigma} - \sin ^2 \theta_W Q_\sigma) (U^\dagger_R)_{i\sigma} (U_R)_{\sigma j},
\label{eq:gZR}
\end{align}
where the electric charge $Q_\sigma$ and the third component of weak isospins $T^3_{L\sigma}$ and $T^3_{R\sigma}$ for each flavor are summarized in Table~\ref{tab:Z_coupling}.
\begin{table}
\centering
\begin{tabular}{l c c c c |}
  \hline
    \hline
Flavor  & 2 & 4 & 5 \\
 \hline
$Q_{\sigma}$ & $-$1 & $-$1 & $-$1 \\ \hline
$T^{3}_{L\sigma}$ & $-$1/2 & $-$1/2 & 0 \\ \hline
$T^{3}_{R\sigma}$ & 0 & $-$1/2 & 0 \\
   \hline
  \hline
      \end{tabular}
\caption{This table shows the electric charge $Q_\sigma$ and the third component of weak isospins $T^3_{L\sigma}$ and $T^3_{R\sigma}$ for each flavor considered in the mixing of charged leptons. Recall that in the flavor basis, we have $2=\mu, 4=L$, and $5 =E$.} 
\label{tab:Z_coupling}
\end{table}

Since the electric charge $Q_\sigma$ is the same for all the charged leptons considered, no modification is made on the couplings of photon with charged leptons. On the other hand, because of the different weak isospins of the charged leptons in the mixing, the couplings of the $Z$ boson in Eq.\,(\ref{eq:gZL}) and Eq.\,(\ref{eq:gZR}) are modified from their SM values. In $(g^{Z}_L)_{ij}$, we have
\begin{align}
-\frac{g}{\mathrm{cos}\,\theta_{W}}\left\{ \left(-\frac{1}{2}+\mathrm{sin}^{2}\theta_{W}\right)[(U^\dagger_L)_{i2} (U_L)_{2j} + (U^\dagger_L)_{i4} (U_L)_{4j}] + \mathrm{sin}^{2}\theta_{W}(U^\dagger_L)_{i5} (U_L)_{5j} \right\}.
\end{align}
To see the modification to the SM coupling, we can arrange and separate Eq.\,(\ref{eq:gZL}) in two parts. The first one is
\begin{align}
-\frac{g}{\mathrm{cos}\,\theta_{W}}\left\{ \left(-\frac{1}{2}+\mathrm{sin}^{2}\theta_{W}\right)[(U^\dagger_L)_{i2} (U_L)_{2j} + (U^\dagger_L)_{i4} (U_L)_{4j}+(U^\dagger_L)_{i5} (U_L)_{5j}]\right\}.
\end{align}
By the unitarity of the matrix $U_{L}$, we have $\sum_{\sigma=2,4,5}(U^\dagger_L)_{i\sigma} (U_L)_{\sigma j} = \delta_{ij}$. This gives the form of the SM coupling of $Z$ boson with the left-handed leptons
\begin{align}
(g^{Z}_{L})^{\mathrm{SM}}_{ij} = -\frac{g}{\mathrm{cos}\,\theta_{W}} \left(-\frac{1}{2}+\mathrm{sin}^{2}\theta_{W}\right)\delta_{ij}.
\end{align}
The second one corresponds to the modification of the SM coupling,
\begin{align}
(\delta g^{Z}_{L})_{ij} = -\frac{g}{2\,\mathrm{cos}\,\theta_{W}}(U^\dagger_L)_{i5} (U_L)_{5j}.
\end{align}
Similarly, we can separate Eq.\,(\ref{eq:gZR}) in parts of the SM coupling
\begin{align}
(g^{Z}_{R})^{\mathrm{SM}}_{ij} = -\frac{g}{\mathrm{cos}\,\theta_W}\sin^{2}\theta_W  \,\delta_{ij}
\label{eq:gRSM}
\end{align}
 and the modification
 \begin{align}
(\delta g^{Z}_{R})_{ij} = \frac{g}{2\mathrm{cos}\,\theta_W}  (U^\dagger_R)_{i4} (U_R)_{4j}.
\label{eq:delgR}
\end{align}

\subsection{$W$ boson couplings}
The couplings of $W$ with leptons are also derived from the kinetic term. Since the charged lepton $E$ is an SU(2)$_{L}$ singlet, it decouples from the interaction with $W$ boson. The kinetic term is given by
\begin{align}
{\cal L}_{\mathrm{kin}} &\supset  -\frac{g}{\sqrt{2}}  \left(  \bar \nu_{\mu} \gamma^\mu \mu_{L} + \bar L_{L}^0 \gamma^\mu L_{L}^- +   \bar L_{R}^0 \gamma^\mu L_{R}^-  \right)W^+_\mu + h.c.  \\
&= -\frac{g}{\sqrt{2}}  \left(  \bar \nu_{2} \gamma^\mu (U_L)_{2j} f_{L j} + \bar \nu_{L4} \gamma^\mu (U_L)_{4j} f_{L j} +   \bar \nu_{R4} \gamma^\mu (U_R)_{4j} f_{R j}  \right)W^+_\mu + h.c.  .
\label{eq:kinW}
\end{align}
This can be written in a more compact form
\begin{align}
{\cal L}_{W} \supset \left[ \bar \nu_{Li} \gamma^\mu (g^{W}_{L})_{ij} f_{Lj} + \bar \nu_{Ri} \gamma^\mu (g^{W}_R)_{ij} f_{Rj} \right] W^+_\mu + h.c.,
\end{align}
and the couplings of $W$ boson with leptons are defined as
 \begin{align}
 (g^{W}_{L})_{2j} =  -\frac{g}{\sqrt{2}}   (U_L)_{2j},  \qquad  (g^{W}_{L})_{4j} =  -\frac{g}{\sqrt{2}}   (U_L)_{4j}, 
\qquad
(g^{W}_{R})_{4j} =  -\frac{g}{\sqrt{2}}   (U_R)_{4j}~,
\label{eq:gWR}
\end{align}
with $ (g^{W}_{R})_{2j} = (g^{W}_{L})_{5j} = (g^{W}_{R})_{5j} = 0$. 

\subsection{Couplings in the electron EDM formulas}

The interaction of the Higgs boson with the charged leptons is
\begin{align}
\bar{f}_{Li} \, \lambda_{ij} \, f_{Rj} \, h + \bar{f}_{Ri} \, \lambda^{\ast}_{ji} \, f_{Lj} \, h = \bar{f}_{i} ( \lambda_{ij}P_{R} + \lambda^{\ast}_{ji}P_{L} )f_{j} \, h,
\end{align}
where $P_{L,R}$ are the projection operators for the left-handed and right-handed components of the leptons.
This can be rearranged in the form of the scalar and pseudoscalar couplings
\begin{align}
\bar{f}_{i} ( g_{s}^{ij} + g_{p}^{ij}\gamma_{5} )f_{j} \, h,
\end{align}
where
\begin{align}
g_{s}^{ij} = \frac{1}{2}(\lambda_{ij}+\lambda^{\ast}_{ji}) ~,~ g_{p}^{ij} = \frac{1}{2}(\lambda_{ij}-\lambda^{\ast}_{ji})
\end{align}
are the scalar and pseudoscalar types of coupling.

The interaction of charged leptons with $Z$ boson can be written as
\begin{align}
\bar f_{i} \gamma^\mu \left( g^{Z,ij}_{v} + g^{Z,ij}_{a}\gamma_{5} \right) f_{j} Z_\mu,
\end{align}
where
\begin{align}
g^{Z,ij}_{v} = \frac{1}{2}\left[(g^{Z}_{L})_{ij}+(g^{Z}_{R})_{ij}\right] ~,~ g^{Z,ij}_{a} = \frac{1}{2}\left[(-g^{Z}_{L})_{ij}+(g^{Z}_{R})_{ij}\right]
\end{align}
are the vector and axial-vector types of coupling.

The electron couplings in Eq.\,(\ref{eq:electron_edm}) are given by
\begin{align}
g^{ee}_s &= -m_{e}/v, \\
g^{\gamma ee}_v &= e, \\
g^{Zee}_v &= -\frac{g}{4\,\mathrm{cos}\,\theta_{W}}\left( -1 + 4\,\mathrm{sin}^{2}\theta_{W} \right), g^{Zee}_a = -\frac{g}{4\,\mathrm{cos}\,\theta_{W}}.
\end{align}

\section{An example for the muon-only coupling}\label{appendix:muon_only}

In the model discussed in our work, we just assume that the vector-like leptons couple only to the second-generation leptons, not to the first- and third-generation ones. In this appendix, we briefly discuss a model where this setup is realized as a consequence of U(1) gauge symmetries. 

As discussed in Refs.~\cite{Araki:2012ip, Asai:2019ciz}, there are two types of lepton flavor-dependent U(1) symmetries which can be gauged and introduced to a model without anomalies. One is the linear combination of the lepton numbers U(1)$_{\alpha_{e}L_{e}+\alpha_{\mu}L_{\mu}-(\alpha_{e}+\alpha_{\mu})L_{\tau}}$ and the other is a linear combination of baryon and lepton numbers U(1)$_{B+\beta_{e}L_{e}+\beta_{\mu}L_{\mu}-(3+\beta_{e}+\beta_{\mu})L_{\tau}}$. As an example, we choose them to be U(1)$_{L_{\mu}-L_{\tau}}$ and U(1)$_{B+3L_{e}-L_{\mu}-5L_{\tau}}$. The charge assignment of the fields is summarized in Table~\ref{tab:appendix_charge}. Note that we introduce three right-handed neutrinos, $N_e, N_\mu, N_\tau$. To break the new U(1) gauge symmetries, we also introduce two scalar fields that are singlet under the SM gauge interactions, $\sigma$ and $\sigma^{\prime}$.  We choose the charges of the vector-like leptons to be the same as the ones of muon so that the interaction structure of coupling solely to the muon in Eq.\,(\ref{eq:Lagrangian_VLL}) can be realized. The charges of the new scalars are chosen for the reason explained in the following.
\begin{table}
\centering
\begin{tabular}{l | c c c c | c c | c c | c c c}
  \hline
    \hline
   & $e$ & $\mu$ & $\tau$ & $H$ & $L_{L,R}$ & $E_{L,R}$ & $\sigma$ & $\sigma^{\prime}$ & $N_{e}$ & $N_{\mu}$ & $N_{\tau}$ \\
 \hline
U(1)$_{L_{\mu}-L_{\tau}}$ & 0 &  1 & $-$1 & 0 & 1 & 1 & 1 & 0 & 0 & 1 & $-$1 \\ \hline
U(1)$_{B+3L_{e}-L_{\mu}-5L_{\tau}}$ & 3 & $-$1 & $-$5 & 0 & $-$1 & $-$1 & 2 & 6 & 3 & $-$1 & $-$5 \\ 
   \hline
  \hline
      \end{tabular}
\caption{Charge assignment of gauged U(1)$_{L_{\mu}-L_{\tau}}$ and U(1)$_{B+3L_{e}-L_{\mu}-5L_{\tau}}$ symmetries to the fields considered in the model in appendix \ref{appendix:muon_only}.} 
\label{tab:appendix_charge}
\end{table}

After the spontaneous symmetry breaking of scalar fields, we obtain a Majorana mass matrix of the right-handed neutrinos in the form of
\begin{align}
\begin{pmatrix}
 \lambda_{ee} \langle \sigma^{\prime\ast} \rangle & \lambda_{e\mu} \langle \sigma^{\ast} \rangle & \lambda_{e\tau} \langle \sigma \rangle \\
 \lambda_{e\mu} \langle \sigma^{\ast} \rangle & 0 & \lambda_{\mu\tau} \langle \sigma^{\prime} \rangle \\
\lambda_{e\tau} \langle \sigma \rangle & \lambda_{\mu\tau} \langle \sigma^{\prime} \rangle & 0
\end{pmatrix},
\end{align}
while the neutrino Dirac mass matrix is diagonal. This implies that after applying the seesaw mechanism~\cite{Minkowski:1977sc,
Yanagida:1980xy, Gell-Mann:1979vob, Mohapatra:1979ia}, we obtain an active neutrino mass matrix with the so-called two-zero minor structure
\cite{Lavoura:2004tu, Lashin:2007dm}. We can then follow the same analysis performed in~\cite{Asai:2017ryy, Asai:2018ocx, Asai:2020qax} to extract the predictions of the sum of three active neutrino masses and the Dirac $CP$-violating phase $\delta$.

\begin{figure}
\centering
\includegraphics[width=2.8in] {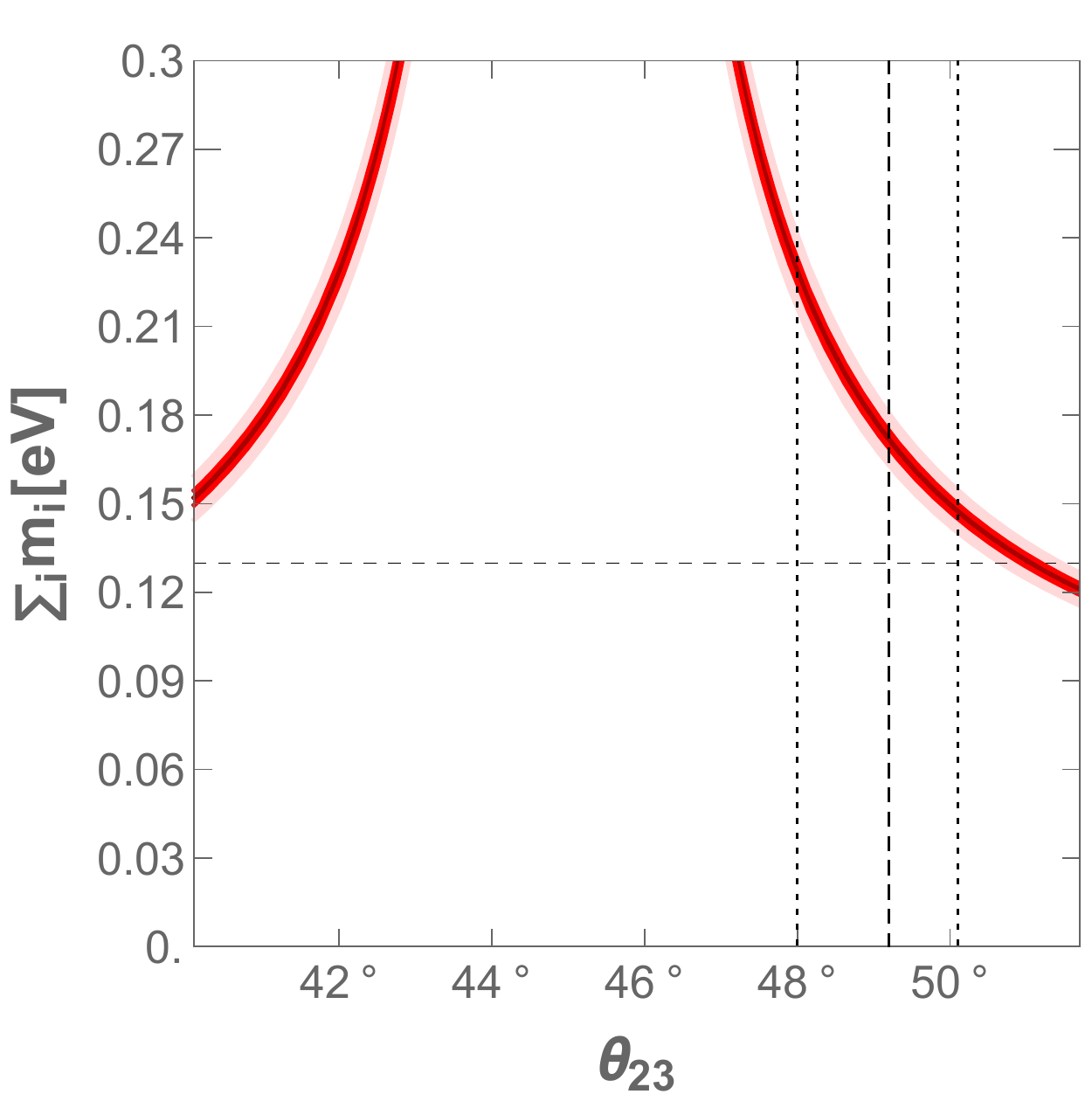}
\hspace{0.5cm}
\includegraphics[width=2.8in] {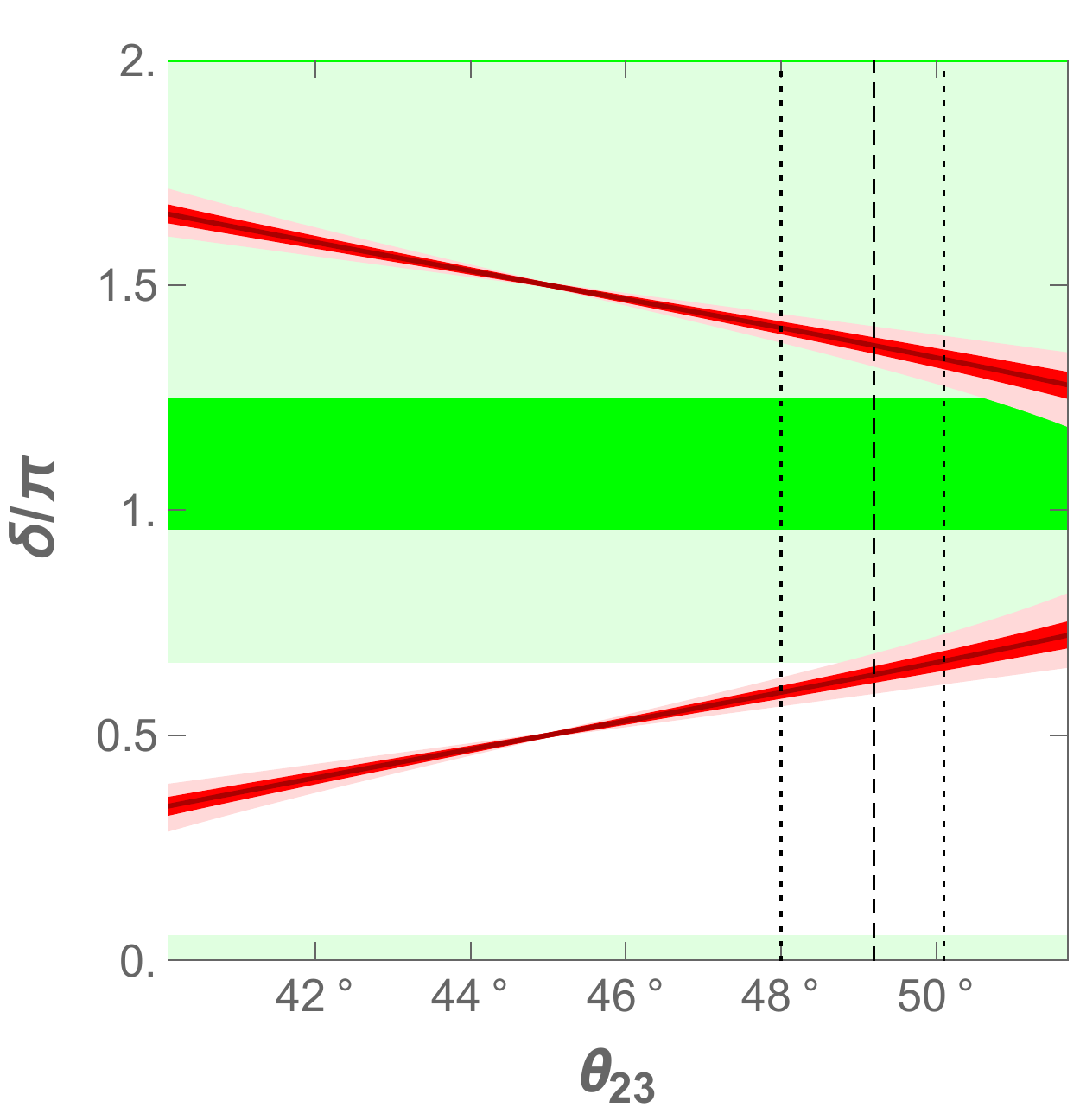}
\caption{Left plot: predictions of the sum of the three active neutrino masses as a function of $\theta_{23}$. Right plot: predictions of the Dirac $CP$ phase as a function of $\theta_{23}$. The predictions are plotted as red curves, dark (light) red bands corresponding to the parameters fixed to their central value and 1 (3) sigma value given in~\cite{Esteban:2020cvm}. In both plots, the boundary of $x$ axis corresponds to the 3$\sigma$ range of $\theta_{23}$, while the vertical dashed line shows the central value of $\theta_{23}$ and the 1$\sigma$ range of $\theta_{23}$ are enclosed by the vertical dotted lines. The horizon dashed line in the left plot represents the upper bound on the sum of the active neutrino masses given in~\cite{DES:2021wwk}, which is $\sum m_{\nu} < 0.13$ eV at 95$\%$ confidence level. The (light) green band in the right plot shows the (3) 1$\sigma$ range of the Dirac $CP$-violating phase obtained in~\cite{Esteban:2020cvm}.}
\label{fig:mu_tau_model}
\end{figure}

In Fig.~\ref{fig:mu_tau_model}, we show these predictions as a function of the neutrino mixing angle $\theta_{23}$, with other two mixing angles and the two squared mass differences fixed by the values provided in~\cite{Esteban:2020cvm}. The neutrino masses are taken to be in the normal ordering, $m_{1} < m_{2} < m_{3}$, as the inverse ordering turns out to be incompatible with the observed neutrino oscillation data in this model~\cite{Asai:2017ryy}. In both plots, the boundary of $x$ axis corresponds to the 3$\sigma$ range of $\theta_{23}$, while the vertical dashed line shows the central value of $\theta_{23}$ and the 1$\sigma$ range of $\theta_{23}$ are enclosed by the vertical dotted lines. The horizon dashed line in the left plot represents the upper bound on the sum of active neutrino masses given in~\cite{DES:2021wwk}, which is $\sum m_{\nu} < 0.13$ eV at 95$\%$ confidence level. The (light) green band in the right plot shows the (3) 1$\sigma$ range of the Dirac $CP$-violating phase obtained in~\cite{Esteban:2020cvm}. The predictions from the two-zero minor structure are plotted as red curves, dark (light) red bands corresponding to the parameters fixed to their central value and 1 (3) sigma value, respectively, given in~\cite{Esteban:2020cvm}. As we can see from the plots, although a large fraction of the parameter region are excluded by the constraint on the sum of the neutrino masses, there is still a possibility around the corner close to the 3$\sigma$ boundary of $\theta_{23}$, which can be examined in the future neutrino experiments.

\bibliographystyle{utphysmod}
\bibliography{ref}

\providecommand{\href}[2]{#2}\begingroup\raggedright\begin{thebibliography}{10}

\bibitem{Muong-2:2021ojo}
{\bfseries Muon g-2} Collaboration, {\em {Measurement of the Positive Muon
  Anomalous Magnetic Moment to 0.46 ppm}},
  \href{https://dx.doi.org/10.1103/PhysRevLett.126.141801}{Phys.\  Rev.\
  Lett.\  {\bfseries 126} (2021) 141801} {\ttfamily
  [\href{https://arxiv.org/abs/2104.03281}{arXiv:2104.03281}]}.

\bibitem{Aoyama:2020ynm}
T.~Aoyama {\em et~al.}, {\em {The anomalous magnetic moment of the muon in the
  Standard Model}},
  \href{https://dx.doi.org/10.1016/j.physrep.2020.07.006}{Phys.\  Rept.\
  {\bfseries 887} (2020) 1--166} {\ttfamily
  [\href{https://arxiv.org/abs/2006.04822}{arXiv:2006.04822}]}.

\bibitem{Borsanyi:2020mff}
S.~Borsanyi {\em et~al.}, {\em {Leading hadronic contribution to the muon
  magnetic moment from lattice QCD}},
  \href{https://dx.doi.org/10.1038/s41586-021-03418-1}{Nature {\bfseries 593}
  (2021) 51--55} {\ttfamily
  [\href{https://arxiv.org/abs/2002.12347}{arXiv:2002.12347}]}.

\bibitem{Ce:2022kxy}
M.~C\`e {\em et~al.}, {\em {Window observable for the hadronic vacuum
  polarization contribution to the muon $g-2$ from lattice QCD}}, {\ttfamily
  \href{https://arxiv.org/abs/2206.06582}{arXiv:2206.06582}} (2022).

\bibitem{Alexandrou:2022amy}
C.~Alexandrou {\em et~al.}, {\em {Lattice calculation of the short and
  intermediate time-distance hadronic vacuum polarization contributions to the
  muon magnetic moment using twisted-mass fermions}}, {\ttfamily
  \href{https://arxiv.org/abs/2206.15084}{arXiv:2206.15084}} (2022).

\bibitem{Pospelov:2005pr}
M.~Pospelov and A.~Ritz, {\em {Electric dipole moments as probes of new
  physics}}, \href{https://dx.doi.org/10.1016/j.aop.2005.04.002}{Annals Phys.\
  {\bfseries 318} (2005) 119--169} {\ttfamily
  [\href{https://arxiv.org/abs/hep-ph/0504231}{hep-ph/0504231}]}.

\bibitem{Muong-2:2008ebm}
{\bfseries Muon (g-2)} Collaboration, {\em {An Improved Limit on the Muon
  Electric Dipole Moment}},
  \href{https://dx.doi.org/10.1103/PhysRevD.80.052008}{Phys.\  Rev.\  D
  {\bfseries 80} (2009) 052008} {\ttfamily
  [\href{https://arxiv.org/abs/0811.1207}{arXiv:0811.1207}]}.

\bibitem{ACME:2018yjb}
{\bfseries ACME} Collaboration, {\em {Improved limit on the electric dipole
  moment of the electron}},
  \href{https://dx.doi.org/10.1038/s41586-018-0599-8}{Nature {\bfseries 562}
  (2018) 355--360}.

\bibitem{Abe:2019thb}
M.~Abe {\em et~al.}, {\em {A New Approach for Measuring the Muon Anomalous
  Magnetic Moment and Electric Dipole Moment}},
  \href{https://dx.doi.org/10.1093/ptep/ptz030}{PTEP {\bfseries 2019} (2019)
  053C02} {\ttfamily
  [\href{https://arxiv.org/abs/1901.03047}{arXiv:1901.03047}]}.

\bibitem{Adelmann:2021udj}
A.~Adelmann {\em et~al.}, {\em {Search for a muon EDM using the frozen-spin
  technique}}, {\ttfamily
  \href{https://arxiv.org/abs/2102.08838}{arXiv:2102.08838}} (2021).

\bibitem{Kannike:2011ng}
K.~Kannike, M.~Raidal, D.~M.~Straub, and A.~Strumia, {\em {Anthropic solution
  to the magnetic muon anomaly: the charged see-saw}},
  \href{https://dx.doi.org/10.1007/JHEP02(2012)106}{JHEP {\bfseries 02} (2012)
  106} {\ttfamily [\href{https://arxiv.org/abs/1111.2551}{arXiv:1111.2551}]}.
  [Erratum: JHEP 10, 136 (2012)].

\bibitem{Falkowski:2013jya}
A.~Falkowski, D.~M.~Straub, and A.~Vicente, {\em {Vector-like leptons: Higgs
  decays and collider phenomenology}},
  \href{https://dx.doi.org/10.1007/JHEP05(2014)092}{JHEP {\bfseries 05} (2014)
  092} {\ttfamily [\href{https://arxiv.org/abs/1312.5329}{arXiv:1312.5329}]}.

\bibitem{Dermisek:2013gta}
R.~Dermisek and A.~Raval, {\em {Explanation of the Muon g-2 Anomaly with
  Vectorlike Leptons and its Implications for Higgs Decays}},
  \href{https://dx.doi.org/10.1103/PhysRevD.88.013017}{Phys.\  Rev.\  D
  {\bfseries 88} (2013) 013017} {\ttfamily
  [\href{https://arxiv.org/abs/1305.3522}{arXiv:1305.3522}]}.

\bibitem{Arnan:2016cpy}
P.~Arnan, L.~Hofer, F.~Mescia, and A.~Crivellin, {\em {Loop effects of heavy
  new scalars and fermions in $b\to s\mu^+\mu^-$}},
  \href{https://dx.doi.org/10.1007/JHEP04(2017)043}{JHEP {\bfseries 04} (2017)
  043} {\ttfamily [\href{https://arxiv.org/abs/1608.07832}{arXiv:1608.07832}]}.

\bibitem{Megias:2017dzd}
E.~Megias, M.~Quiros, and L.~Salas, {\em {$g_\mu-2$ from Vector-Like Leptons in
  Warped Space}}, \href{https://dx.doi.org/10.1007/JHEP05(2017)016}{JHEP
  {\bfseries 05} (2017) 016} {\ttfamily
  [\href{https://arxiv.org/abs/1701.05072}{arXiv:1701.05072}]}.

\bibitem{Kowalska:2017iqv}
K.~Kowalska and E.~M.~Sessolo, {\em {Expectations for the muon g-2 in
  simplified models with dark matter}},
  \href{https://dx.doi.org/10.1007/JHEP09(2017)112}{JHEP {\bfseries 09} (2017)
  112} {\ttfamily [\href{https://arxiv.org/abs/1707.00753}{arXiv:1707.00753}]}.

\bibitem{Raby:2017igl}
S.~Raby and A.~Trautner, {\em {Vectorlike chiral fourth family to explain muon
  anomalies}}, \href{https://dx.doi.org/10.1103/PhysRevD.97.095006}{Phys.\
  Rev.\  D {\bfseries 97} (2018) 095006} {\ttfamily
  [\href{https://arxiv.org/abs/1712.09360}{arXiv:1712.09360}]}.

\bibitem{Poh:2017tfo}
Z.~Poh and S.~Raby, {\em {Vectorlike leptons: Muon g-2 anomaly, lepton flavor
  violation, Higgs boson decays, and lepton nonuniversality}},
  \href{https://dx.doi.org/10.1103/PhysRevD.96.015032}{Phys.\  Rev.\  D
  {\bfseries 96} (2017) 015032} {\ttfamily
  [\href{https://arxiv.org/abs/1705.07007}{arXiv:1705.07007}]}.

\bibitem{Kawamura:2019rth}
J.~Kawamura, S.~Raby, and A.~Trautner, {\em {Complete vectorlike fourth family
  and new U(1)' for muon anomalies}},
  \href{https://dx.doi.org/10.1103/PhysRevD.100.055030}{Phys.\  Rev.\  D
  {\bfseries 100} (2019) 055030} {\ttfamily
  [\href{https://arxiv.org/abs/1906.11297}{arXiv:1906.11297}]}.

\bibitem{Hiller:2019mou}
G.~Hiller, C.~Hormigos-Feliu, D.~F.~Litim, and T.~Steudtner, {\em {Anomalous
  magnetic moments from asymptotic safety}},
  \href{https://dx.doi.org/10.1103/PhysRevD.102.071901}{Phys.\  Rev.\  D
  {\bfseries 102} (2020) 071901} {\ttfamily
  [\href{https://arxiv.org/abs/1910.14062}{arXiv:1910.14062}]}.

\bibitem{Hiller:2020fbu}
G.~Hiller, C.~Hormigos-Feliu, D.~F.~Litim, and T.~Steudtner, {\em {Model
  Building from Asymptotic Safety with Higgs and Flavor Portals}},
  \href{https://dx.doi.org/10.1103/PhysRevD.102.095023}{Phys.\  Rev.\  D
  {\bfseries 102} (2020) 095023} {\ttfamily
  [\href{https://arxiv.org/abs/2008.08606}{arXiv:2008.08606}]}.

\bibitem{Endo:2020tkb}
M.~Endo and S.~Mishima, {\em {Muon $g-2$ and CKM unitarity in extra lepton
  models}}, \href{https://dx.doi.org/10.1007/JHEP08(2020)004}{JHEP {\bfseries
  08} (2020) 004} {\ttfamily
  [\href{https://arxiv.org/abs/2005.03933}{arXiv:2005.03933}]}.

\bibitem{Frank:2020smf}
M.~Frank and I.~Saha, {\em {Muon anomalous magnetic moment in two-Higgs-doublet
  models with vectorlike leptons}},
  \href{https://dx.doi.org/10.1103/PhysRevD.102.115034}{Phys.\  Rev.\  D
  {\bfseries 102} (2020) 115034} {\ttfamily
  [\href{https://arxiv.org/abs/2008.11909}{arXiv:2008.11909}]}.

\bibitem{Chun:2020uzw}
E.~J.~Chun and T.~Mondal, {\em {Explaining $g-2$ anomalies in two Higgs doublet
  model with vector-like leptons}},
  \href{https://dx.doi.org/10.1007/JHEP11(2020)077}{JHEP {\bfseries 11} (2020)
  077} {\ttfamily [\href{https://arxiv.org/abs/2009.08314}{arXiv:2009.08314}]}.

\bibitem{Kowalska:2020zve}
K.~Kowalska and E.~M.~Sessolo, {\em {Minimal models for g-2 and dark matter
  confront asymptotic safety}},
  \href{https://dx.doi.org/10.1103/PhysRevD.103.115032}{Phys.\  Rev.\  D
  {\bfseries 103} (2021) 115032} {\ttfamily
  [\href{https://arxiv.org/abs/2012.15200}{arXiv:2012.15200}]}.

\bibitem{Dermisek:2021ajd}
R.~Dermisek, K.~Hermanek, and N.~McGinnis, {\em {Muon g-2 in two-Higgs-doublet
  models with vectorlike leptons}},
  \href{https://dx.doi.org/10.1103/PhysRevD.104.055033}{Phys.\  Rev.\  D
  {\bfseries 104} (2021) 055033} {\ttfamily
  [\href{https://arxiv.org/abs/2103.05645}{arXiv:2103.05645}]}.

\bibitem{Lee:2022nqz}
H.~M.~Lee and K.~Yamashita, {\em {A model of vector-like leptons for the muon
  $g-2$ and the W boson mass}},
  \href{https://dx.doi.org/10.1140/epjc/s10052-022-10635-z}{Eur.\  Phys.\  J.\
  C {\bfseries 82} (2022) 661} {\ttfamily
  [\href{https://arxiv.org/abs/2204.05024}{arXiv:2204.05024}]}.

\bibitem{deGiorgi:2022xhr}
A.~de~Giorgi, L.~Merlo, and S.~Pokorski, {\em {The Low-Scale Seesaw Solution to
  the $M_W$ and $(g-2)_\mu$ Anomalies}}, {\ttfamily
  \href{https://arxiv.org/abs/2211.03797}{arXiv:2211.03797}} (2022).

\bibitem{Babu:2000dq}
K.~S.~Babu, B.~Dutta, and R.~N.~Mohapatra, {\em {Enhanced electric dipole
  moment of the muon in the presence of large neutrino mixing}},
  \href{https://dx.doi.org/10.1103/PhysRevLett.85.5064}{Phys.\  Rev.\  Lett.\
  {\bfseries 85} (2000) 5064--5067} {\ttfamily
  [\href{https://arxiv.org/abs/hep-ph/0006329}{hep-ph/0006329}]}.

\bibitem{Ibrahim:2001jz}
T.~Ibrahim and P.~Nath, {\em {Slepton flavor nonuniversality, the muon EDM and
  its proposed sensitive search at Brookhaven}},
  \href{https://dx.doi.org/10.1103/PhysRevD.64.093002}{Phys.\  Rev.\  D
  {\bfseries 64} (2001) 093002} {\ttfamily
  [\href{https://arxiv.org/abs/hep-ph/0105025}{hep-ph/0105025}]}.

\bibitem{Feng:2001sq}
J.~L.~Feng, K.~T.~Matchev, and Y.~Shadmi, {\em {Theoretical expectations for
  the muon's electric dipole moment}},
  \href{https://dx.doi.org/10.1016/S0550-3213(01)00383-2}{Nucl.\  Phys.\  B
  {\bfseries 613} (2001) 366--381} {\ttfamily
  [\href{https://arxiv.org/abs/hep-ph/0107182}{hep-ph/0107182}]}.

\bibitem{Romanino:2001zf}
A.~Romanino and A.~Strumia, {\em {Electron and Muon Electric Dipoles in
  Supersymmetric Scenarios}},
  \href{https://dx.doi.org/10.1016/S0550-3213(01)00607-1}{Nucl.\  Phys.\  B
  {\bfseries 622} (2002) 73--94} {\ttfamily
  [\href{https://arxiv.org/abs/hep-ph/0108275}{hep-ph/0108275}]}.

\bibitem{Ellis:2001xt}
J.~R.~Ellis, J.~Hisano, S.~Lola, and M.~Raidal, {\em {CP violation in the
  minimal supersymmetric seesaw model}},
  \href{https://dx.doi.org/10.1016/S0550-3213(01)00583-1}{Nucl.\  Phys.\  B
  {\bfseries 621} (2002) 208--234} {\ttfamily
  [\href{https://arxiv.org/abs/hep-ph/0109125}{hep-ph/0109125}]}.

\bibitem{Ellis:2001yza}
J.~R.~Ellis, J.~Hisano, M.~Raidal, and Y.~Shimizu, {\em {Lepton electric dipole
  moments in nondegenerate supersymmetric seesaw models}},
  \href{https://dx.doi.org/10.1016/S0370-2693(02)01197-8}{Phys.\  Lett.\  B
  {\bfseries 528} (2002) 86--96} {\ttfamily
  [\href{https://arxiv.org/abs/hep-ph/0111324}{hep-ph/0111324}]}.

\bibitem{Bartl:2003ju}
A.~Bartl, W.~Majerotto, W.~Porod, and D.~Wyler, {\em {Effect of supersymmetric
  phases on lepton dipole moments and rare lepton decays}},
  \href{https://dx.doi.org/10.1103/PhysRevD.68.053005}{Phys.\  Rev.\  D
  {\bfseries 68} (2003) 053005} {\ttfamily
  [\href{https://arxiv.org/abs/hep-ph/0306050}{hep-ph/0306050}]}.

\bibitem{Cheung:2009fc}
K.~Cheung, O.~C.~W.~Kong, and J.~S.~Lee, {\em {Electric and anomalous magnetic
  dipole moments of the muon in the MSSM}},
  \href{https://dx.doi.org/10.1088/1126-6708/2009/06/020}{JHEP {\bfseries 06}
  (2009) 020} {\ttfamily
  [\href{https://arxiv.org/abs/0904.4352}{arXiv:0904.4352}]}.

\bibitem{Hiller:2010ib}
G.~Hiller, K.~Huitu, T.~Ruppell, and J.~Laamanen, {\em {A Large Muon Electric
  Dipole Moment from Flavor?}},
  \href{https://dx.doi.org/10.1103/PhysRevD.82.093015}{Phys.\  Rev.\  D
  {\bfseries 82} (2010) 093015} {\ttfamily
  [\href{https://arxiv.org/abs/1008.5091}{arXiv:1008.5091}]}.

\bibitem{Cesarotti:2018huy}
C.~Cesarotti, Q.~Lu, Y.~Nakai, A.~Parikh, and M.~Reece, {\em {Interpreting the
  Electron EDM Constraint}},
  \href{https://dx.doi.org/10.1007/JHEP05(2019)059}{JHEP {\bfseries 05} (2019)
  059} {\ttfamily [\href{https://arxiv.org/abs/1810.07736}{arXiv:1810.07736}]}.

\bibitem{Dekens:2018bci}
W.~Dekens, J.~de~Vries, M.~Jung, and K.~K.~Vos, {\em {The phenomenology of
  electric dipole moments in models of scalar leptoquarks}},
  \href{https://dx.doi.org/10.1007/JHEP01(2019)069}{JHEP {\bfseries 01} (2019)
  069} {\ttfamily [\href{https://arxiv.org/abs/1809.09114}{arXiv:1809.09114}]}.

\bibitem{Crivellin:2018qmi}
A.~Crivellin, M.~Hoferichter, and P.~Schmidt-Wellenburg, {\em {Combined
  explanations of $(g-2)_{\mu,e}$ and implications for a large muon EDM}},
  \href{https://dx.doi.org/10.1103/PhysRevD.98.113002}{Phys.\  Rev.\  D
  {\bfseries 98} (2018) 113002} {\ttfamily
  [\href{https://arxiv.org/abs/1807.11484}{arXiv:1807.11484}]}.

\bibitem{Altmannshofer:2020ywf}
W.~Altmannshofer, S.~Gori, H.~H.~Patel, S.~Profumo, and D.~Tuckler, {\em
  {Electric dipole moments in a leptoquark scenario for the $B$-physics
  anomalies}}, \href{https://dx.doi.org/10.1007/JHEP05(2020)069}{JHEP
  {\bfseries 05} (2020) 069} {\ttfamily
  [\href{https://arxiv.org/abs/2002.01400}{arXiv:2002.01400}]}.

\bibitem{Bigaran:2021kmn}
I.~Bigaran and R.~R.~Volkas, {\em {Reflecting on chirality: CP-violating
  extensions of the single scalar-leptoquark solutions for the
  (g-2)e,\ensuremath{\mu} puzzles and their implications for lepton EDMs}},
  \href{https://dx.doi.org/10.1103/PhysRevD.105.015002}{Phys.\  Rev.\  D
  {\bfseries 105} (2022) 015002} {\ttfamily
  [\href{https://arxiv.org/abs/2110.03707}{arXiv:2110.03707}]}.

\bibitem{Omura:2015xcg}
Y.~Omura, E.~Senaha, and K.~Tobe, {\em {$\tau$- and $\mu$-physics in a general
  two Higgs doublet model with $\mu-\tau$ flavor violation}},
  \href{https://dx.doi.org/10.1103/PhysRevD.94.055019}{Phys.\  Rev.\  D
  {\bfseries 94} (2016) 055019} {\ttfamily
  [\href{https://arxiv.org/abs/1511.08880}{arXiv:1511.08880}]}.

\bibitem{Hou:2021zqq}
W.-S.~Hou, G.~Kumar, and S.~Teunissen, {\em {Charged lepton EDM with extra
  Yukawa couplings}}, \href{https://dx.doi.org/10.1007/JHEP01(2022)092}{JHEP
  {\bfseries 01} (2022) 092} {\ttfamily
  [\href{https://arxiv.org/abs/2109.08936}{arXiv:2109.08936}]}.

\bibitem{Nakai:2022vgp}
Y.~Nakai, R.~Sato, and Y.~Shigekami, {\em {Muon electric dipole moment as a
  probe of flavor-diagonal CP violation}},
  \href{https://dx.doi.org/10.1016/j.physletb.2022.137194}{Phys.\  Lett.\  B
  {\bfseries 831} (2022) 137194} {\ttfamily
  [\href{https://arxiv.org/abs/2204.03183}{arXiv:2204.03183}]}.

\bibitem{Dermisek:2022aec}
R.~Dermisek, K.~Hermanek, N.~McGinnis, and S.~Yoon, {\em {Ellipse of Muon
  Dipole Moments}},
  \href{https://dx.doi.org/10.1103/PhysRevLett.129.221801}{Phys.\  Rev.\
  Lett.\  {\bfseries 129} (2022) 221801} {\ttfamily
  [\href{https://arxiv.org/abs/2205.14243}{arXiv:2205.14243}]}.

\bibitem{Ema:2021jds}
Y.~Ema, T.~Gao, and M.~Pospelov, {\em {Improved Indirect Limits on Muon
  Electric Dipole Moment}},
  \href{https://dx.doi.org/10.1103/PhysRevLett.128.131803}{Phys.\  Rev.\
  Lett.\  {\bfseries 128} (2022) 131803} {\ttfamily
  [\href{https://arxiv.org/abs/2108.05398}{arXiv:2108.05398}]}.

\bibitem{Barr:1990vd}
S.~M.~Barr and A.~Zee, {\em {Electric Dipole Moment of the Electron and of the
  Neutron}}, \href{https://dx.doi.org/10.1103/PhysRevLett.65.21}{Phys.\  Rev.\
  Lett.\  {\bfseries 65} (1990) 21--24}. [Erratum: Phys.Rev.Lett. 65, 2920
  (1990)].

\bibitem{Nakai:2016atk}
Y.~Nakai and M.~Reece, {\em {Electric Dipole Moments in Natural
  Supersymmetry}}, \href{https://dx.doi.org/10.1007/JHEP08(2017)031}{JHEP
  {\bfseries 08} (2017) 031} {\ttfamily
  [\href{https://arxiv.org/abs/1612.08090}{arXiv:1612.08090}]}.

\bibitem{CMS:2020xwi}
{\bfseries CMS} Collaboration, {\em {Evidence for Higgs boson decay to a pair
  of muons}}, \href{https://dx.doi.org/10.1007/JHEP01(2021)148}{JHEP {\bfseries
  01} (2021) 148} {\ttfamily
  [\href{https://arxiv.org/abs/2009.04363}{arXiv:2009.04363}]}.

\bibitem{LHCHiggsCrossSectionWorkingGroup:2016ypw}
{\bfseries LHC Higgs Cross Section Working Group} Collaboration, {\em {Handbook
  of LHC Higgs Cross Sections: 4. Deciphering the Nature of the Higgs Sector}},
  {\ttfamily \href{https://arxiv.org/abs/1610.07922}{arXiv:1610.07922}} (2016).

\bibitem{Workman:2022}
{\bfseries Particle Data Group} Collaboration, {\em {Review of Particle
  Physics}},. to be published (2022).

\bibitem{L3:2001xsz}
{\bfseries L3} Collaboration, {\em {Search for heavy neutral and charged
  leptons in $e^{+} e^{-}$ annihilation at LEP}},
  \href{https://dx.doi.org/10.1016/S0370-2693(01)01005-X}{Phys.\  Lett.\  B
  {\bfseries 517} (2001) 75--85} {\ttfamily
  [\href{https://arxiv.org/abs/hep-ex/0107015}{hep-ex/0107015}]}.

\bibitem{CMS:2022nty}
{\bfseries CMS} Collaboration, {\em {Inclusive nonresonant multilepton probes
  of new phenomena at $\sqrt s$=13\,\,TeV}},
  \href{https://dx.doi.org/10.1103/PhysRevD.105.112007}{Phys.\  Rev.\  D
  {\bfseries 105} (2022) 112007} {\ttfamily
  [\href{https://arxiv.org/abs/2202.08676}{arXiv:2202.08676}]}.

\bibitem{ACMEexp}
{\em {The ACME EDM Experiment}}, \url{http://doylegroup.harvard.edu/edm/}.

\bibitem{Alarcon:2022ero}
R.~Alarcon {\em et~al.} in {\em {2022 Snowmass Summer Study}}.
\newblock 2022.
\newblock {\ttfamily
  \href{https://arxiv.org/abs/2203.08103}{arXiv:2203.08103}}.

\bibitem{Araki:2012ip}
T.~Araki, J.~Heeck, and J.~Kubo, {\em {Vanishing Minors in the Neutrino Mass
  Matrix from Abelian Gauge Symmetries}},
  \href{https://dx.doi.org/10.1007/JHEP07(2012)083}{JHEP {\bfseries 07} (2012)
  083} {\ttfamily [\href{https://arxiv.org/abs/1203.4951}{arXiv:1203.4951}]}.

\bibitem{Asai:2019ciz}
K.~Asai, {\em {Predictions for the neutrino parameters in the minimal model
  extended by linear combination of U(1)$_{L_e-L_\mu}$, U(1)$_{L_\mu-L_\tau}$
  and U(1)$_{B-L}$ gauge symmetries}},
  \href{https://dx.doi.org/10.1140/epjc/s10052-020-7622-6}{Eur.\  Phys.\  J.\
  C {\bfseries 80} (2020) 76} {\ttfamily
  [\href{https://arxiv.org/abs/1907.04042}{arXiv:1907.04042}]}.

\bibitem{Minkowski:1977sc}
P.~Minkowski, {\em {$\mu \to e\gamma$ at a Rate of One Out of $10^{9}$ Muon
  Decays?}}, \href{https://dx.doi.org/10.1016/0370-2693(77)90435-X}{Phys.\
  Lett.\  B {\bfseries 67} (1977) 421--428}.

\bibitem{Yanagida:1980xy}
T.~Yanagida, {\em {Horizontal Symmetry and Masses of Neutrinos}},
  \href{https://dx.doi.org/10.1143/PTP.64.1103}{Prog.\  Theor.\  Phys.\
  {\bfseries 64} (1980) 1103}.

\bibitem{Gell-Mann:1979vob}
M.~Gell-Mann, P.~Ramond, and R.~Slansky, {\em {Complex Spinors and Unified
  Theories}}, Conf.\  Proc.\  C {\bfseries 790927} (1979) 315--321 {\ttfamily
  [\href{https://arxiv.org/abs/1306.4669}{arXiv:1306.4669}]}.

\bibitem{Mohapatra:1979ia}
R.~N.~Mohapatra and G.~Senjanovic, {\em {Neutrino Mass and Spontaneous Parity
  Nonconservation}},
  \href{https://dx.doi.org/10.1103/PhysRevLett.44.912}{Phys.\  Rev.\  Lett.\
  {\bfseries 44} (1980) 912}.

\bibitem{Lavoura:2004tu}
L.~Lavoura, {\em {Zeros of the inverted neutrino mass matrix}},
  \href{https://dx.doi.org/10.1016/j.physletb.2005.01.047}{Phys.\  Lett.\  B
  {\bfseries 609} (2005) 317--322} {\ttfamily
  [\href{https://arxiv.org/abs/hep-ph/0411232}{hep-ph/0411232}]}.

\bibitem{Lashin:2007dm}
E.~I.~Lashin and N.~Chamoun, {\em {Zero minors of the neutrino mass matrix}},
  \href{https://dx.doi.org/10.1103/PhysRevD.78.073002}{Phys.\  Rev.\  D
  {\bfseries 78} (2008) 073002} {\ttfamily
  [\href{https://arxiv.org/abs/0708.2423}{arXiv:0708.2423}]}.

\bibitem{Asai:2017ryy}
K.~Asai, K.~Hamaguchi, and N.~Nagata, {\em {Predictions for the neutrino
  parameters in the minimal gauged U(1)$_{L_\mu-L_\tau}$ model}},
  \href{https://dx.doi.org/10.1140/epjc/s10052-017-5348-x}{Eur.\  Phys.\  J.\
  C {\bfseries 77} (2017) 763} {\ttfamily
  [\href{https://arxiv.org/abs/1705.00419}{arXiv:1705.00419}]}.

\bibitem{Asai:2018ocx}
K.~Asai, K.~Hamaguchi, N.~Nagata, S.-Y.~Tseng, and K.~Tsumura, {\em {Minimal
  Gauged U(1)$_{L_\alpha - L_\beta}$ Models Driven into a Corner}},
  \href{https://dx.doi.org/10.1103/PhysRevD.99.055029}{Phys.\  Rev.\  D
  {\bfseries 99} (2019) 055029} {\ttfamily
  [\href{https://arxiv.org/abs/1811.07571}{arXiv:1811.07571}]}.

\bibitem{Asai:2020qax}
K.~Asai, K.~Hamaguchi, N.~Nagata, and S.-Y.~Tseng, {\em {Leptogenesis in the
  minimal gauged U(1)$_{L_\mu-L_\tau}$ model and the sign of the cosmological
  baryon asymmetry}},
  \href{https://dx.doi.org/10.1088/1475-7516/2020/11/013}{JCAP {\bfseries 11}
  (2020) 013} {\ttfamily
  [\href{https://arxiv.org/abs/2005.01039}{arXiv:2005.01039}]}.

\bibitem{Esteban:2020cvm}
I.~Esteban, M.~C.~Gonzalez-Garcia, M.~Maltoni, T.~Schwetz, and A.~Zhou, {\em
  {The fate of hints: updated global analysis of three-flavor neutrino
  oscillations}}, \href{https://dx.doi.org/10.1007/JHEP09(2020)178}{JHEP
  {\bfseries 09} (2020) 178} {\ttfamily
  [\href{https://arxiv.org/abs/2007.14792}{arXiv:2007.14792}]}.

\bibitem{DES:2021wwk}
{\bfseries DES} Collaboration, {\em {Dark Energy Survey Year 3 results:
  Cosmological constraints from galaxy clustering and weak lensing}},
  \href{https://dx.doi.org/10.1103/PhysRevD.105.023520}{Phys.\  Rev.\  D
  {\bfseries 105} (2022) 023520} {\ttfamily
  [\href{https://arxiv.org/abs/2105.13549}{arXiv:2105.13549}]}.

\end{thebibliography}\endgroup


\providecommand{\href}[2]{#2}\begingroup\raggedright\endgroup


\end{document}